\documentclass[aps,prl,reprint,amsmath,amssymb,superscriptaddress, 10pt]{revtex4-2}

\usepackage[usenames,dvipsnames]{xcolor}
\usepackage{amsmath,amssymb}
\usepackage{graphicx}
\usepackage{hyperref}
\usepackage{braket}
\usepackage[normalem]{ulem}
\usepackage{enumerate}
\usepackage{MnSymbol}
\usepackage{esint}

\newcommand{\be}{\begin{equation}}
\newcommand{\ee}{\end{equation}}
\def\bes{\begin{subequations}}
\def\esu{\end{subequations}}

\newcommand{\dr}{\text{dr}}

\newcommand{\eff}{\text{eff}}

\newcommand{\mg}{\nu}

\usepackage{etoolbox}
\usepackage{orcidlink}

\newcommand{\up}{{\uparrow}}
\newcommand{\dw}{{\downarrow}}

\newcommand{\com}[1]{{}} 

\newcommand{\dd}{{\rm d}}

\newcommand{\para}[1]{ \noindent\emph{\textbf{#1---}}}

\begin{document}

\newcommand{\titleinfo}{Generalized Hydrodynamics of Bloch Oscillations in the Absence of a Lattice}

\title{\titleinfo}

\author{Stefano Scopa~\orcidlink{00-0001-7638-8804}}
\affiliation{Laboratoire de Physique de l’\'Ecole Normale Superieure, CNRS,
ENS \& Universit\'e PSL, Sorbonne Universit\'e, Universit\'e Paris Cit\'e, 75005 Paris, France.}

\author{Philip Zechmann~\orcidlink{0000-0002-9714-2480}}
\affiliation{Technical University of Munich, TUM School of Natural Sciences, Physics Department, 85748 Garching, Germany}
\affiliation{Munich Center for Quantum Science and Technology (MCQST), Schellingstr. 4, 80799 M{\"u}nchen, Germany}

\author{Michael Knap~\orcidlink{0000-0001-7877-0329}}
\affiliation{Technical University of Munich, TUM School of Natural Sciences, Physics Department, 85748 Garching, Germany}
\affiliation{Munich Center for Quantum Science and Technology (MCQST), Schellingstr. 4, 80799 M{\"u}nchen, Germany}

\author{Jacopo De Nardis~\orcidlink{0000-0001-7877-0329}}
\affiliation{Laboratoire de Physique Th\'eorique et Mod\'elisation, CNRS UMR 8089, CY Cergy Paris Universit\'e, 95302 Cergy-Pontoise Cedex, France.}
\affiliation{JEIP, UAR 3573 CNRS, Collège de France, PSL Research University, 11 Place Marcelin Berthelot, 75321 Paris Cedex 05, France.}

\author{Alvise Bastianello~\orcidlink{0000-0003-3441-671X}}
\affiliation{CEREMADE, CNRS, Universit\'e Paris-Dauphine, Universit\'e PSL, 75016 Paris, France}
\email{alvise.bastianello@dauphine.psl.eu}

\begin{abstract}
Objects subjected to a constant force generally increase their velocity over time. This expectation fails whenever their energy is a smooth and periodic function of momentum, resulting in periodic Bloch oscillations instead. Periodic dispersions, typical of lattice systems, can also emerge in continuum media through strong interactions.
Here, we study the phenomenon of  such Bloch oscillations in the absence of a lattice in a paradigmatic model of integrable quantum gases: the two-component Yang-Gaudin model. 
We derive a generalized-hydrodynamic theory of Bloch oscillations for a finite density of impurities embedded in a homogeneous interacting background, which we show to persist superimposed to a drift due to the acceleration of the center of mass.
Moreover, we show the single-impurity oscillation period is renormalized at finite impurity density when two-magnon bound states are populated.
Our results are relevant for ultracold atom experiments, where impurities can be created at controllable densities.
\end{abstract}

\maketitle

\para{\textbf{Introduction.}} 
Everyday experience suggests that a force acting on an object increases its velocity, converting potential energy into kinetic energy.
In this naive expectation, the dispersion law is crucial: for example, relativistic particles cannot overcome the speed of light. Even richer are lattice systems, where the kinetic energy is typically a bounded and periodic function of momentum, giving rise to persistent Bloch oscillations~\cite{bloch1929,zener1934,Dahan1996,Gustavsson2008}.

\begin{figure}[b!]
\centering
	\includegraphics[width=0.8\columnwidth]{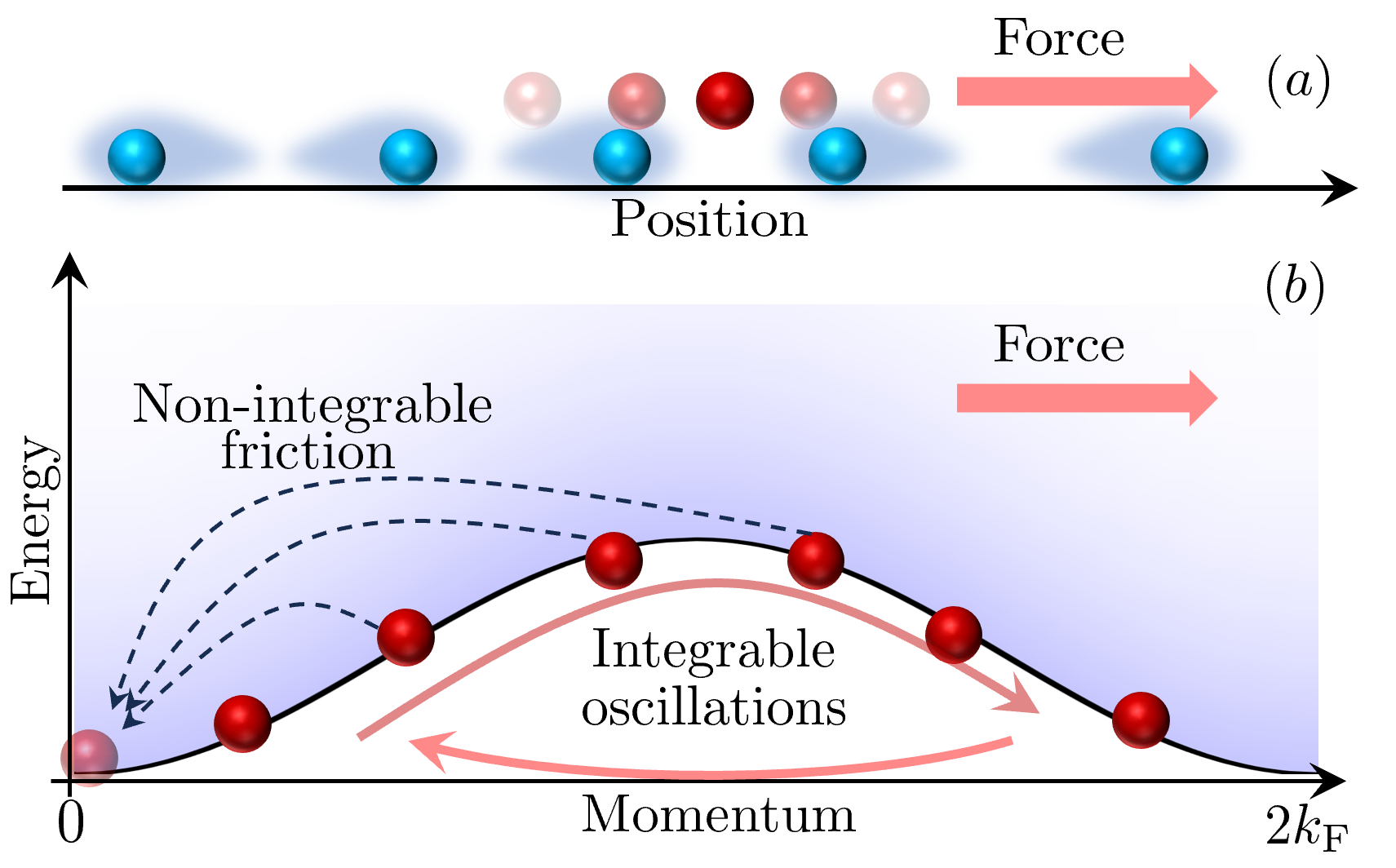}
\caption{\textbf{Phenomenology of Bloch oscillations.---} Panel (a): An impurity interacting with a background gas undergoes periodic Bloch oscillations in real space when accelerated with a constant force. Panel (b): The single-impurity dispersion on top of the ground state is a periodic function of momentum. In the non-integrable case, the impurity experiences friction due to inelastic collisions with the continuum spectrum (purple shading), whereas friction is suppressed in the integrable case.
Hydrodynamic Bloch oscillations persit at finite impurity density within generalized hydrodynamics.
}
	\label{fig_1}
\end{figure}

In strongly interacting quantum systems, Bloch oscillations can arise as an \emph{emergent} phenomenon even in the continuum. Impurities form polarons~\cite{Schmidt2018}, new quasiparticles dressed by interactions with the background, whose properties can be profoundly different from those of the bare particles. Hence, the polaron dispersion law becomes periodic, see Fig.~\ref{fig_1}.
Bloch oscillations without a lattice arise in ultracold one-dimensional gases, where the universal properties of the ground state are captured by the Tomonaga-Luttinger liquid~\cite{giamarchi2003quantum} and an emergent Fermi sea with Fermi momentum $k_\text{F}=\pi n$, with $n$ the density of the gas. Phenomenologically, an impurity experiences an effective lattice set by the scale $2 k_\text{F}$, hence undergoing Bloch oscillations with period $\Delta t=2\hbar k_\text{F}/F$ when accelerated with a constant force of strength $F$.
Following the seminal proposal~\cite{Gangardt2009}, extensive work~\cite{Matveev2008,Kamenev2009,Schecter2012,Schecter2012,Gamayun2014,Schecter2015,Gamayun2015,Yang2017,Lychkovskiy2018,Majumdar2026} has characterized this phenomenon, culminating in its experimental realization~\cite{Meinert2017,Rabec2025}.

Exact Bloch oscillations are generally expected in the quantum adiabatic limit~\cite{Berry2009}, where the applied force is so small that the polaron follows its dispersion law. However, since in the thermodynamic limit there is no gap between the polaron's energy and a continuum of states, accelerating impurities radiate energy into the host gas and lead to a finite drift velocity, with possibly superimposed oscillations~\cite{Gangardt2009,Schecter2012,Gamayun2014,Schecter2015,Gamayun2015}.
Finite-temperature effects also cause friction through inelastic scattering with thermal phonons in the background gas, leading to oscillations with amplitude $\propto F^2$~\cite{Gangardt2009}, see Fig.~\ref{fig_1}(b).

Given the multitude of corrections involved, it appears hard to observe many oscillations in experiments~\cite{Meinert2017}.
However, a particular class of systems may offer a more robust realization of Bloch oscillations: these are nearly integrable systems~\cite{takahashi2005thermodynamics},
featuring infinitely many conservation laws and stable quasiparticle excitations that scatter elastically. The absence of inelastic scattering with the background hampers conventional friction of the impurity with the background~\cite{Gangardt2009}, see Fig.~\ref{fig_1}(b).

At the many-body level, the absence of inelastic scatterings prevents thermalization in favor of more general stationary states called generalized Gibbs ensembles (GGE)~\cite{Rigol2009,Langen2015,Calabrese2016}. 
Even in the absence of energy gaps, an uncontrolled proliferation of excitations is prevented by the additional conservation laws. Nearly integrable models describe a multitude of ultracold atom experiments~\cite{Guan2022} and greatly contributed to our understanding of non-equilibrium quantum physics. In particular, generalized hydrodynamics (GHD)~\cite{Alvaredo2016,Bertini2016,Bastianello2022,Doyon2024}---a non-perturbative hydrodynamic approach to integrable systems---has enabled quantitative comparisons between experiments and theory~\cite{Schemmer2019,Malvania2021,Schuttelkopf2024,dubois2024,Cataldini2022,Moller2021,Yang2024Phantom,horvath2025,Zeng2026}. It is natural to apply GHD to the phenomenon of Bloch oscillations in the absence of a lattice.

In this Letter, we consider the integrable Yang-Gaudin (YG) model~\cite{Yang1967}, which describes a two-component gas of bosons or fermions with two-body contact interactions.
Motivated by experimental progress~\cite{Guan2013}, we here consider the fermionic YG model whose GHD has been previously applied in Refs.~\cite{Mestyan2019,Scopa2021,Scopa2022} to study spin-charge separation~\cite{giamarchi2003quantum}.
Using GHD to describe weak forces, we analytically determine the emergence of \emph{hydrodynamic} Bloch oscillations even at finite impurity density and temperature, superimposed on an uniform acceleration of the center of mass. Generalizing the single-impurity phenomenology, we unveil a periodic evolution of the instantaneous many-body GGE. Interactions renormalize the period away from $2\hbar k_\text{F}/F$ whenever two-magnon bound states are excited. Our findings are supported by analytically solvable limits of the GHD equations and by numerical simulations beyond these cases.

\para{The model.} We consider the Hamiltonian
\be\label{eq_YG_H}
\hat{H}=\frac{\hbar^2}{2m}\int \!\dd x \bigg( \partial_x\hat{\psi}_\up^\dagger\partial_x\hat{\psi}_\up+\partial_x\hat{\psi}_\dw^\dagger\partial_x\hat{\psi}_\dw+\frac{2mc}{\hbar^2} \hat{\psi}^\dagger_\up\hat{\psi}_\up\hat{\psi}^\dagger_\dw\hat{\psi}_\dw\bigg) ,
\ee
where $\hat{\psi}_{\up,\dw}$ are canonical fermionic fields. The system is in the repulsive regime $c>0$ and initially prepared in a thermal state with a density imbalance for the two species, $n_\up>n_\dw$. 
Hence, the species $\up$ describes the host gas, whereas we dub the species $\dw$ as the ``impurity". We notice that the impurity is non-generic, in the sense that its interactions with the host component preserve integrability, whereas in the general case impurities break integrability~\cite{Schecter2012}.
Hereafter, $n=n_\up+n_\dw$ is the total density.
The system is let to evolve under the effect of a force coupled to the impurity, $\hat{H}\to \hat{H}_F=\hat{H}-\int \dd x\, F x \, \hat{\psi}^\dagger_\dw\hat{\psi}_\dw$.
For small force terms, such that particles relax through several scattering events before the force induces sizeable effects, the system hydrodynamically evolves through stationary states of the unperturbed Hamiltonian.

\para{Conventional hydrodynamics.} We first consider the predictions of conventional hydrodynamics (CHD), to be later contrasted with GHD.
Hence, for the time being, instead of Eq. \eqref{eq_YG_H} we consider a non-integrable Galilean invariant Hamiltonian, which is then perturbed by the force. CHD assumes only four conserved quantities: the number of particles of both species, the energy, and the momentum.
Due to Galilean invariance, the current operator $\hat{j}=\hat{j}_\up+\hat{j}_\dw$, with $\hat{j}_{\up,\dw}=-\,m^{-1}\left(i\hat{\psi}_{\up,\dw}^\dagger\partial_x\hat{\psi}_{\up,\dw}+\text{h.c.}\right)$, evolves as $\partial_t \hat{j}=\frac{1}{m}F \hat{\psi}^\dagger_\dw\hat{\psi}_\dw$, while the unperturbed part of the Hamiltonian obeys $\partial_t\hat{H}=F \int\dd x\,\hat{j}_\dw$. The center-of-mass velocity is defined as $v_\text{cm}=\langle \hat{j}\rangle/n$ and it linearly grows in time, $v_\text{cm}=F t m^{-1}n_\dw/n$.
To access the impurity current, we invoke the CHD assumption $\langle j_\dw\rangle_\text{CHD}=v_\text{cm}n_\dw$, from which the energy is determined as $\langle \hat{H}\rangle_{\text{CHD}}=\text{const}+\frac{1}{2}m N v_\text{cm}^2$, with $N$ the total number of particles. Hence, the initial state is constantly accelerated with no heating in the boosted reference frame, nor Bloch oscillations. Physically, the impurity undergoes inelastic scattering with bulk excitations, transferring momentum and experiencing friction~\cite{Gangardt2009,Schecter2012,Gamayun2014,Schecter2015,Gamayun2015}: in the CHD regime, friction dominates over the small force $F$, and the impurity average momentum relaxes to that of the host gas.

\para{Generalized hydrodynamics.}
GHD properly accounts for the infinite hierarchy of conserved quantities of the YG Hamiltonian~\eqref{eq_YG_H}~\cite{Scopa2021,Scopa2022}.
GHD is best formulated in terms of the emergent quasiparticles, parametrized by a real number $\lambda\in \mathbb{R}$ called ``rapidity'', generalizing the wavevector of non-interacting systems, and a ``string index'' $j$, spanning over different species of excitations. 
In the YG model, two classes of quasiparticles coexist: the string $j=0$ describes the matter degrees of freedom regardless of their spin, whereas strings $j\ge 1$ capture spin excitations. In particular, the $j^\text{th}$ quasiparticle describes a magnonic bound state of $j$ spins. We define the magnetization function $\mg_j(\lambda)=j$, and denote as $\epsilon_j(\lambda)$ and $p_j(\lambda)$ the bare energy and momentum of each quasiparticle species. In the YG model, $\epsilon_0(\lambda)=\tfrac{\hbar^2}{2m}\lambda^2$ and $p_0(\lambda)=\hbar\lambda$, while spin degrees of freedom do not carry bare energy nor momentum, $\epsilon_{j>0}(\lambda)=p_{j>0}(\lambda)=0$.
The system evolves according to the GHD equation~\cite{Doyon2017,Bastianello2019}
\be\label{eq_GHD}
\partial_t\rho_{j;t}(\lambda)+\partial_\lambda[a^\eff_{j;t}(\lambda)\rho_{j;t}(\lambda)]=0\, .
\ee
We focus on the homogeneous case, but generalizations to inhomogeneous settings are possible. Hereafter, we drop the explicit $t-$dependence to ease the notation.
Above, $\rho_{j}(\lambda)$ is the density of excitations associated with the instantaneous GGE, and the effective acceleration $a^\eff_{j}(\lambda)$ is renormalized by interactions as $a^\eff_{j}(\lambda)=F[ \mg_{j}(\lambda)]^\dr/[\partial_\lambda p_{j}(\lambda)]^\dr$, where the dressing operation for an arbitrary test function $\tau_j(\lambda)\to [\tau_j(\lambda)]^\dr$ is defined as
\be\label{eq_dressing}
[\tau_j(\lambda)]^\dr=\tau_j(\lambda)-\int \frac{\dd\lambda'}{2\pi} \sum_{j'}\Phi_{j,j'}(\lambda-\lambda')\vartheta_{j'}(\lambda')[\tau_{j'}(\lambda')]^\dr\, ,
\ee
where the occupancy, or filling function, is $\vartheta_j(\lambda)=2\pi \hbar \rho_j(\lambda)/[\partial_\lambda p_j(\lambda)]^\dr$. Notice that $a^\eff_j(\lambda)$ is a highly nonlinear function of the state itself, and thus it evolves in time. The scattering shift $\Phi_{j,j'}(\lambda)$ describes the space displacement after a scattering event due to interactions~\cite{Doyon2018,Doyon2024TT,Urilyon2026}. It is determined by the elementary function $\phi_j(\lambda)=\tfrac{\dd}{\dd\lambda}2\arctan\left[\frac{2\hbar^2 \lambda}{mc j}\right]$ as $\Phi_{0,j>0}(\lambda)=\Phi_{j>0,0}(\lambda)=-\phi_2(\lambda)$, $\Phi_{j>0,j'>0}(\lambda)=(1-\delta_{j,j'})\phi_{|j-j'|}(\lambda)+2\phi_{|j-j'|+2}(\lambda)+...+2\phi_{j+j'-2}(\lambda)+\phi_{j+j'}(\lambda)$, and $\Phi_{0,0}(\lambda)=0$.
Within GHD, the particle densities are given by $n=\int \dd\lambda \rho_0(\lambda)$ and $n_\dw=\sum_j\int \dd\lambda \mg_j(\lambda) \rho_j(\lambda)$, while the associated currents are $\langle \hat{j}\rangle_\text{GHD}=\int \dd\lambda  v^\eff_0(\lambda)\rho_0(\lambda)$ and $\langle \hat{j}_\dw\rangle_\text{GHD}=\sum_j\int \dd\lambda  v^\eff_j(\lambda)\mg_j(\lambda)\rho_j(\lambda)$. The effective velocity $v^\eff_j(\lambda)$ is defined as $v^\eff_j(\lambda)=[\partial_\lambda \epsilon_j(\lambda)]^\dr/[\partial_\lambda p_j(\lambda)]^\dr$~\cite{Alvaredo2016,Bertini2016}.
Equation~\eqref{eq_GHD} gives $\langle \hat{j}\rangle_{\text{GHD}}=tm^{-1}F n_\dw$, as it should be by Galilean invariance; see the End Matter.
The nonlinear equation~\eqref{eq_GHD} is analytically solved in certain limits, which we now analyze before numerically addressing the general case.

\begin{figure}[t!]
\centering
	\includegraphics[width=0.99\columnwidth]{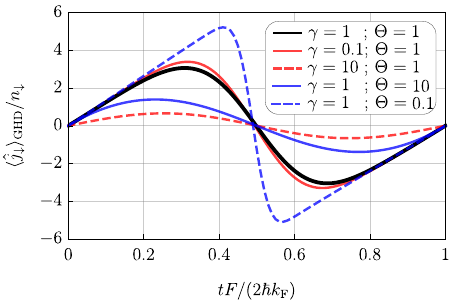}
\caption{\textbf{Bloch oscillations for low impurity density.---} We show the normalized impurity current $\langle \hat{j}_\dw\rangle_\text{GHD}/n_\dw$ generated by the effect of the external force $F$, for different values of the dimensionless interaction $\gamma$ and temperature $\Theta$. We consider $\gamma=1$ and $\Theta=1$ (black) as a reference, and vary alternatively the interaction (red) and the temperature (blue).}
	\label{fig_2}
\end{figure}

\para{The low impurity-density limit.} We analytically solve Eq.~\eqref{eq_GHD} for $n_\dw\ll n_\up$, where $\rho_{j>0}(\lambda),\vartheta_{j>0}(\lambda)\to 0$, and establish the emergence of Bloch oscillations \emph{for any temperature}. 
Equation~\eqref{eq_GHD} can be conveniently reparametrized as $\partial_t\vartheta_j+a^\eff_j\partial_\lambda \vartheta_j=0$~\cite{Alvaredo2016,Bertini2016,Doyon2017,Bastianello2019}, which describes an inhomogeneous translation of $\vartheta_j$ in the rapidity space.  
At leading order in $\vartheta_{j>0}(\lambda)$, Eq.~\eqref{eq_dressing} gives $a^\eff_j(\lambda)\simeq F\mg_j/[\partial_\lambda p_j(\lambda)]^\dr$ and $[\partial_\lambda p_{j>0}(\lambda)]^\dr\simeq -\int \frac{\dd\lambda'}{2\pi} \Phi_{j,0}(\lambda-\lambda')\bar{\rho}_0(\lambda')$, where $\bar{\rho}_0(\lambda)$ is the background constant excitation density. On thermal states, it has a Fermi-Dirac distribution $\bar{\rho}_0(\lambda)=\frac{1}{2\pi}[1+e^{\beta (\epsilon_0(\lambda)-\mu)}]^{-1}$, with $\beta$ the inverse temperature and $\mu$ the chemical potential.
Since $\bar{\rho}_0(\lambda)$ has finite support, $a_j^\eff(\lambda)$ diverges for large $|\lambda|$, accelerating $\vartheta_{j>0}(\lambda)$ to infinite rapidity in finite time. This apparent singularity is resolved by introducing, for each string, a more convenient parametrization, $k(\lambda)=\int_{-\infty}^\lambda\dd\lambda' [\partial_\lambda p_j(\lambda')]^\dr$. Notice that the new variable lives in a Brillouin zone with $k(+\infty)=2\hbar k_\text{F}$.

\begin{figure}[t!]
\centering
	\includegraphics[width=0.99\columnwidth]{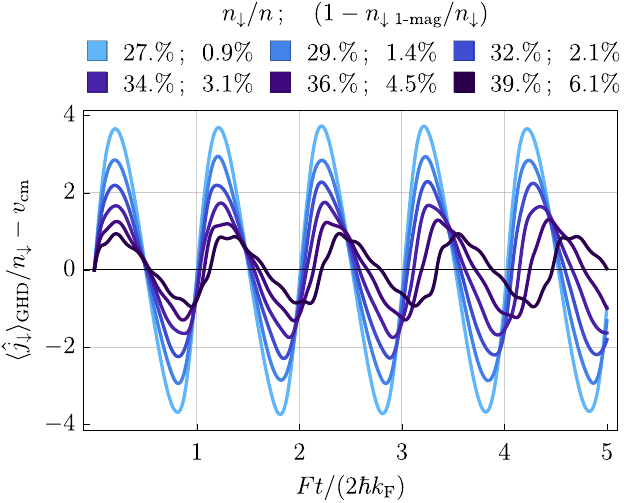}
\caption{\textbf{The weakly interacting limit.---} We show the normalized impurity current in the center-of-mass frame $\langle \hat{j}_\dw\rangle_\text{GHD}/n_\dw-v_\text{cm}$ for $\gamma\to 0$, starting from thermal states with $\Theta=1$, for different impurity densities $n_\dw$. Bloch oscillations are perfect in the absence of excitations beyond one-magnon quasiparticles, and spoiled in their presence: we also report the relative impurity density carried by these degrees of freedom, $(1-n_{\dw\text{1-mag}}/n_\dw)$. }
	\label{fig_3}
\end{figure}
\begin{figure*}[t!]
\centering
	\includegraphics[width=0.99\textwidth]{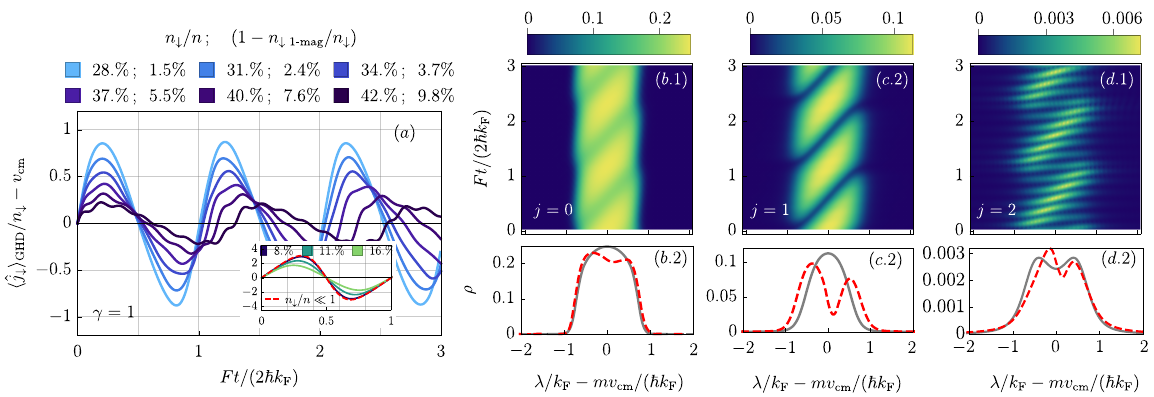}
\caption{\textbf{Bloch oscillations at finite interaction and impurity density.---} We consider initial thermal states with fixed temperature $\Theta=1$ and interaction $\gamma=1$, for different values of the impurity density.
Panel (a) parallels Fig.~\ref{fig_3}, showing the impurity current for different values of $n_\dw/n$. For sufficiently large impurity density, two-magnon bound states are excited and appreciably renormalize the observed oscillation period. Inset: for small impurity density, curves converge to the zero-density impurity limit (dashed red line).
In the right panels we follow the evolution of $\rho_j$ corresponding to $n_\dw/n=0.4$: panels (b.1), (c.1), and (d.1) show the evolution of the strings $j=0$, $j=1$, and $j=2$, respectively. Panels (b.2), (c.2), and (d.2) focus on cuts of the density profiles, comparing the initial profile (gray solid lines) with their evolution at $Ft/(2\hbar k_\text{F})=3$ (dashed red lines), and show their deviation from perfect Bloch oscillations due to the two-magnon excitations.}
	\label{fig_4}
\end{figure*}

Defining $\tilde{\vartheta}_j(k(\lambda))\equiv\vartheta_j(\lambda)$, the leading-order GHD equation becomes $\partial_t\tilde{\vartheta}_j+ F \mg_j\partial_k \tilde{\vartheta}_j=0$, where periodic boundary conditions must be imposed (see also Ref.~\cite{BastianelloFlux2019}). The function $\tilde{\vartheta}_j$ rigidly translates in $k$ space with a period $2\hbar k_\text{F}/(Fj)$: the longest period, obtained for $j=1$, dictates the period of Bloch oscillations in physical observables. In this limit, the center of mass does not accelerate. Notice that this result applies to any physical quasiparticle density $\bar{\rho}_0(\lambda)$, and thus describes ground states, finite-temperature states, and more exotic GGEs.
Figure~\ref{fig_2} shows the impurity current for different values of the dimensionless interaction $\gamma=2mc/(\hbar^2n)$ and temperature $\Theta= k_\text{B} T 2m/(\hbar^2 n^2)$, which modify the shape and amplitude of Bloch oscillations, but neither their existence nor period. Thermal states are determined through standard thermodynamic Bethe ansatz~\cite{takahashi2005thermodynamics}; see End Matter. At a fixed temperature, the amplitude of the Bloch oscillations is reduced upon increasing interactions $\gamma$. At fixed $\gamma$, the amplitude increases by reducing the temperature, approaching a straight line suddenly reflected through Umklapp scattering at the edges of the Brillouin zone.

\para{Finite impurity density: analytically tractable limits.} We now go beyond the infinitesimal impurity limit. Equation~\eqref{eq_GHD} can be analytically solved in two cases: the impenetrable $(c\to\infty)$ and the weakly-interacting $(c\to 0)$ regimes. For impenetrable particles, the amplitude of the Bloch oscillations is found to vanish and the impurity current has an exact linear drift $\langle j_\dw\rangle_\text{GHD}=n_\dw v_\text{cm}$, as in the CHD case; see End Matter.

In contrast, the more interesting weakly interacting limit shows hydrodynamic Bloch oscillations.
It should be emphasized that the hydrodynamics of the non-interacting gas $c=0$, which does not feature Bloch oscillations, is very different from considering the interacting hydrodynamics in the limit $c\to0$, due to the latter having additional conservation laws. In other words, the limit $c\to 0$ does not commute with the hydrodynamic one.
From Fermi Golden-Rule arguments~\cite{Furst2013,Bertini2015,Surace2023}, the crossover time $\tau_c$ between the non-interacting and YG hydrodynamics 
is dictated by second-order perturbation theory in the interaction terms, and thus scaling as $\tau_c\propto \hbar E_\text{F}/(c^2 n_\up n_\dw)$, where $E_\text{F}=\frac{\hbar^2 k_\text{F}^2}{2m}$ is the Fermi energy. Consistently, for the YG hydrodynamics to emerge, the period of the Bloch oscillations must be much larger than the crossover scale $2\hbar k_\text{F}/F\gg \tau_c$, leading to the condition $F\ll  c^2 n_\up n_\dw 4m/(\hbar^2 k_\text{F})$. We assume to work in this limit, and treat it as a useful exercise to better understand the finite-interaction case later on. 
Equation~\eqref{eq_dressing} simplifies to algebraic equations since $\lim_{c\to 0}\phi_n(\lambda)=2\pi \delta(\lambda)$. 
As a minimal yet illustrative example, we truncate the hierarchy of occupancies to the matter occupancy $\vartheta_{0}(\lambda)$, and to the one-magnon $\vartheta_{1}(\lambda)$ and two-magnon $\vartheta_{2}(\lambda)$ states. This is justified for not too large impurity densities and low temperatures: higher truncations can be considered as well; see End Matter.
The validity of this truncation can be determined numerically solving the thermodynamic Bethe ansatz equations and checking the population of higher magnon bound states; see End matter.

In this subspace, the GHD equation is greatly simplified after a change of variable similar in spirit to the one previously considered for a small impurity density: we define $y(\lambda)=2\pi\hbar\int_{-\infty}^\lambda\dd\lambda' \rho_0(\lambda')$. Notice that, as $\rho_0$ evolves in time, the function $y(\lambda)$ is also time-dependent, but $y(+\infty)=2\hbar k_\text{F}$. By defining $\tilde{\vartheta}_{j}(y(\lambda))=\vartheta_{j}(\lambda)$ and using the chain rule $\partial_t\vartheta_{j}(\lambda)=\partial_t\tilde{\vartheta}_{j}(y(\lambda))+\partial_t y(\lambda) \partial_y\tilde{\vartheta}_{j}(y(\lambda))$, the GHD equations become $\partial_t\tilde{\vartheta}_{j}(y)+(a^\eff_j-a^\eff_0)2\pi\hbar\rho_0\partial_y\tilde{\vartheta}_{j}(y)=0$. After some algebraic manipulations, one obtains $\partial_t\tilde{\vartheta}_{0}(y)=0$ and
\be\label{eq_weak}
\partial_t\tilde{\vartheta}_{1}(y)+F\frac{1-\tilde{\vartheta}_2}{1+\tilde{\vartheta}_2}\partial_y\tilde{\vartheta}_1=0\, ,\,\,\,
\partial_t\tilde{\vartheta}_2(y)+F\frac{2}{1-\tilde{\vartheta}_1}\partial_y\tilde{\vartheta}_2=0\, ,
\ee
see End Matter for a detailed derivation.
In the new variable, the matter degrees of freedom are static, whereas the spin degrees of freedom oscillate. If two-magnon bound states are not excited, $\tilde{\vartheta}_2=0$, then the one-magnon occupancy $\tilde{\vartheta}_{1}$ undergoes \emph{perfect} Bloch oscillations even at finite impurity density and temperature, with period $\Delta t=2\hbar k_\text{F}/F$. These perfect Bloch oscillations are spoiled when two-magnon bound states are activated. These excitations contribute with higher frequencies to the oscillations and renormalize the overall period compared with the zero-density impurity limit; see Fig.~\ref{fig_3}. We do not observe a net current as we move in the center-of-mass reference frame.

\para{Finite interactions and impurity density.}
We resort to numerical solutions of the full GHD equation~\eqref{eq_GHD} to explore the generic regime. The gas is initialized in thermal ensembles for a representative choice of interaction, $\gamma=1$ and temperature, $\Theta=1$, and we change the relative impurity density $n_\dw/n$. Other parameters led to qualitatively similar results. The GHD equations are solved with the method of characteristics~\cite{Bastianello2022,Moller2020} after a proper discretization of the rapidity space~\cite{suppmat}. 
Figure~\ref{fig_4}(a) shows the impurity current after subtracting the overall center-of-mass drift. In the weakly interacting limit, we showed that magnonic bound states are responsible for renormalizing the oscillation period: surprisingly, this persists even at finite interactions. This feature can be highlighted by artificially suppressing strings $j> 1$; see SM~\cite{suppmat}. Figures~\ref{fig_4}(b)--(d) follow a representative evolution of the excitation densities $\rho_j(\lambda)$, showing that the quasiperiodic evolution a property of the whole state and highlighting the corrections to perfect Bloch oscillations for finite density of impurity.

\para{Conclusions.} In continuous systems, interactions between the host gas and impurities are ultimately at the origin of Bloch oscillations in the absence of a microscopic lattice. While great attention has been dedicated to the single-impurity problem, in this Letter we unveiled the existence of hydrodynamic Bloch oscillations in integrable systems, which feature an infinite hierarchy of conservation laws accounted for by generalized hydrodynamics. The persistence of the phenomenon at finite temperature and impurity density paves the way to its observation in ultracold atoms.
The key experimental challenge is accessing the hydrodynamic regime through small forces and large times. While precisely characterizing the emergence of the hydrodynamic scale is an open theoretical question, the success of GHD in describing ultracold gases in the presence of trapping potentials and integrability breaking~\cite{Schemmer2019,Malvania2021,Schuttelkopf2024,dubois2024,Cataldini2022,Moller2021,Yang2024Phantom,horvath2025,Zeng2026} shows the hydrodynamic regime is within experimental capabilities.

For example, ideal platforms to observe our results are ${}^6\text{Li}$ atoms, which offer a faithful implementation of the fermionic YG model with spin-resolved detection~\cite{Senaratne2022}. Future studies of real-space manifestations of Bloch oscillations, like correlation functions and the propagation of spatially localized impurities, will provide quantitative experimental predictions.
Generalizing our results to the bosonic YG model~\cite{Klauser2011,Robinson2017} will further broaden reachable experimental realizations. For example, Bloch oscillations of magnetic solitons in ${}^{87}\text{Rb}$ atoms---close to the YG regime---have been recently observed~\cite{Rabec2025}, and understood by theoretically modeling the spin dynamics within the Landau-Lifshitz equation while assuming frozen density of atoms~\cite{Kosevich1998,Kosevich2001,Congy2016,Bresolin2023}. Applying our methodology to the bosonic YG model will capture fluctuations far from this idealized regime. Further directions include exploring the Manakov model~\cite{manakov1974}, viewed as the semiclassical limit~\cite{Koch2022,Koch2023,Bastianello2024} of the YG model, which well describes both weakly interacting cold atoms~\cite{Lannig2020,Cominotti2023,Rabec2025} and photonic platforms~\cite{Menyuk1989,kivshar2003}.

\para{Data and information availability.} Data and simulation codes are available on Zenodo~\cite{Zenodo}.

\para{Acknowledgements.} 
We thank Dimitri Gangardt and Mikhail Zvonarev for useful discussions.
AB is supported by the ERC StG 101220476–NonErgHydro, SS by the MSCA 101103348–GENESYS. 
PZ and MK acknowledge support from the Deutsche Forschungsgemeinschaft (DFG, German Research Foundation) under Germany’s Excellence Strategy–EXC–2111–390814868, TRR 360 – 492547816 and DFG grants No. KN1254/1-2, KN1254/2-1, the European Union (grant agreement No 101169765), as well as the Munich Quantum Valley, which is supported by the Bavarian state government with funds from the Hightech Agenda Bayern Plus. JDN is funded by the ERC StG 101042293–HEPIQ and the ANR-22-CPJ1-0021-01. 
Views and opinions expressed are those of the authors only and do not necessarily reflect those of the European Union or the European Research Council Executive Agency. Neither the European Union nor the granting authority can be held responsible for them.

\bigskip\bigskip

\begin{center}
{\Large \textbf{End Matter}}
\end{center}

\subsection*{Thermodynamic Bethe ansatz}

The occupancy $\vartheta_j(\lambda)$ in thermal Gibbs ensembles is determined by standard methods of thermodynamic Bethe ansatz~\cite{takahashi2005thermodynamics}. Consider a thermal state with density matrix  $\hat{\rho}=\frac{1}{\mathcal{Z}}\exp\left[-\beta \hat{H}_\text{YG}-\beta\mu_\up \hat{N}_\up-\beta\mu_\dw\hat{N}_\dw\right]$ where $\beta$ is the inverse temperature, $\hat{N}_{\up,\dw}$ are the number operators of the two species and $\mu_{\up,\dw}$ the associated chemical potentials. The occupancies associated with Gibbs  states are determined through the non-linear equations
\begin{multline}\label{eq_tba}
\varepsilon_j(\lambda)=\beta (\epsilon_j(\lambda)+\mu_\up\delta_{j,0})+\beta (\mu_\dw-\mu_\up)\nu_j(\lambda)\\
+\sum_{j'}\int \frac{\dd\lambda'}{2\pi} \Phi_{j,j'}(\lambda-\lambda')\log\left(1+e^{\varepsilon_{j'}(\lambda')}\right)\, ,
\end{multline}
where the pseudoenergies $\varepsilon_j(\lambda)$ parametrize the filling functions as $\vartheta_j(\lambda)=[1+e^{\varepsilon_j(\lambda)}]^{-1}$. These equations can be efficiently discretized and numerically solved, see SM~\cite{suppmat}.
In the limit of vanishing impurity density, obtained for $\mu_\dw-\mu_\up\to +\infty$, Eq.~\eqref{eq_tba} can be analytically solved giving $\varepsilon_{j=0}(\lambda)\simeq\beta (\epsilon_j(\lambda)+\mu_\up\delta_{j,0})$ and $\varepsilon_{j\ge 0}(\lambda)\simeq \beta (\mu_\dw-\mu_\up)\nu_j(\lambda)
+\sum_{j'}\int \frac{\dd\lambda'}{2\pi} \Phi_{j,0}(\lambda-\lambda')\log\left(1+e^{-\varepsilon_{0}(\lambda')}\right)$. We used this approximated solution in Fig.~\ref{fig_2}.

\subsection*{The acceleration of the center of mass in GHD}

Due to Galilean invariance, the center of mass must linearly accelerate in time as $v_\text{cm}=Ftm^{-1}n_\dw/n$. We consistently check this prediction is satisfied within GHD. We consider the total density current $\langle \hat{j}\rangle_\text{GHD}$: since the YG model is Galilean invariant, the particle current is proportional to the total momentum and we thus alternatively write $\langle \hat{j}\rangle_\text{GHD}=m^{-1}\int \dd\lambda \, p_0(\lambda) \rho_0$ and take the time derivative $\partial_t\langle \hat{j}\rangle_\text{GHD}=m^{-1}\int \dd\lambda\,  p_0(\lambda)\partial_\lambda(-a^\eff_0 \rho_0)=m^{-1}\int \dd\lambda\, \partial_\lambda p_0(\lambda) a^\eff_0 \rho_0$, where we used the GHD equations and integrated by parts. We now conveniently sum over the strings using that $p_j(\lambda)$ vanishes for $j\ge 0$. By expliciting the effective acceleration we have
\be
\partial_t\langle \hat{j}\rangle_\text{GHD}=Fm^{-1}\sum_j\int \frac{\dd\lambda}{2\pi\hbar} (\partial_\lambda p_j(\lambda)) [\nu_j(\lambda)]^\dr \vartheta_j(\lambda)\, .
\ee
Finally, we use the symmetry of the dressing equation $\sum_j\int \dd\lambda [A_j(\lambda)]^\dr B_j(\lambda)\vartheta_j(\lambda)=\sum_j\int \dd\lambda A_j(\lambda) [B_j(\lambda)]^\dr\vartheta_j(\lambda)$, which holds for every test functions $A_j(\lambda)$, $B_j(\lambda)$. Since it holds $\sum_j\int \frac{\dd\lambda}{2\pi \hbar} [\partial_\lambda p_j(\lambda)]^\dr \nu_j(\lambda) \vartheta_j(\lambda)=n_\dw$, one immediately gets $\partial_t\langle \hat{j}\rangle_\text{GHD}=Fm^{-1} n_\dw$, which does not depend on the evolution of the state, but only on the impurity density. Hence, the impurity current grows with a constant rate and we consistently obtain $v_\text{cm}=Ftm^{-1}n_\dw/n$.

\subsection*{The impenetrable limit in GHD}

As the infinitely repulsive regime is approached $c\to+\infty$, the impurities are not able to scatter through the host particles any longer, and the momentum gain induced by the force is immediately transferred to the whole gas. Following this heuristic argument, one expects the whole system to rigidly accelerate following the velocity of the center of mass.
This expectation is recovered within GHD. Firstly, we observe that the scattering kernel vanishes as $1/c$, hence Eq. \eqref{eq_dressing} can be approximately solved replacing $[\tau_j(\lambda)]^{\rm dr}\to \tau_j(\lambda)$ on the right hand side. This immediately gives that the mass string effective velocity reduces to the bare group velocity $v^\text{eff}_0(\lambda)\simeq v(\lambda)=\hbar\lambda/m$. We now address the effective velocity of the magnetic degrees of freedom.

We notice that  $c$ sets the typical scale of variations of the scattering kernel $\Phi_{j,j'}(\lambda)$ as $\lambda\sim 1/c$ while, at the same time, the support of $\rho_0(\lambda)$ remains finite and determined by the temperature. Within these approximations, one gets $[\partial_\lambda p_{j>0}(\lambda)]^\dr\simeq -\Phi_{j,0}(\lambda) n_\text{tot}$ and $[\partial_\lambda \epsilon_{j>0}(\lambda)]^\dr\simeq - \Phi_{j,0}(\lambda) n_\text{tot}v_\text{cm}$. Therefore, we find a constant effective velocity for the spin degrees of freedom $v_{j>0}^\eff(\lambda)=v_\text{cm}$.
Eq.~\eqref{eq_GHD} is still non-trivial in the spin degrees of freedom, with $\vartheta_{j>0}$ traveling across the whole Brillouin zone. However, since the spin effective velocity is constant, this evolution is undetectable in the impurity current $\langle \hat{j}_\dw\rangle_\text{GHD}=n_\dw v_\text{cm}$ and other degrees of freedom: the amplitude of Bloch oscillations vanishes in the impenetrable limit.

\subsection*{The weakly interacting limit in GHD}
Here we provide further details on the GHD equations in the limit $c\to 0$: we focus on states featuring at most two-magnon bound states, truncating the state to the string $j=2$, and thus considering only $\{\vartheta_0(\lambda),\vartheta_1(\lambda),\vartheta_2(\lambda)\}$. Higher truncations can be numerically considered following the same strategy, but they lead to more cumbersome expressions.
Since in the $c\to 0$ limit the kernel $\Phi_{j,j'}(\lambda)$ approaches a Dirac $\delta$ in the rapidity space, the dressing equation becomes an algebraic equation diagonal in the rapidity space. In particular, the dressing $\tau\to\tau^\dr$ becomes
\be
\begin{pmatrix}
\tau_0^\dr\\ \tau_1^\dr \\ \tau_2^\dr
\end{pmatrix}=\begin{pmatrix}
\tau_0\\ \tau_1 \\ \tau_2
\end{pmatrix}-\begin{pmatrix} 0&& -1 && -1\\ -1 && 1 && 2  \\ -1 && 2 && 3\end{pmatrix}\begin{pmatrix} \vartheta_0 && 0 && 0\\ 0 && \vartheta_1 && 0\\ 0 && 0&& \vartheta_2\end{pmatrix}\begin{pmatrix}
\tau_0^\dr\\ \tau_1^\dr\\ \tau_1^\dr
\end{pmatrix}\, ,
\ee
where we omitted the rapidity variable.
The effective acceleration is then found as
\be
\begin{pmatrix}
a^\eff_0\\ a^\eff_1\\a^\eff_2
\end{pmatrix}=F\begin{pmatrix}
\frac{\vartheta_1+2\vartheta_2-\vartheta_1\vartheta_2}{1+\vartheta_1+3\vartheta_2-\vartheta_1\vartheta_2}\\[6pt]
\frac{1-\vartheta_2+\vartheta_0\vartheta_2}{\vartheta_0+\vartheta_0\vartheta_2}\\[6pt]
\frac{2-\vartheta_0\vartheta_1}{\vartheta_0-\vartheta_0\vartheta_1}
\end{pmatrix}\, .
\ee
The expression is cumbersome, but it greatly simplifies by changing coordinates as
\be\label{eq_ydef}
y(\lambda)=2\pi\hbar \int_{-\infty}^\lambda \dd\lambda'\, \rho_0(\lambda')\, .
\ee
Notice that, although the change of variable is time-dependent, it always holds $y(+\infty)=2 \hbar k_\text{F}$.
Where $\rho_0$ can be determined as a function of the fillings through the dressing
\be
2\pi\hbar  \rho_0= \frac{\vartheta_0(1+\vartheta_1+3\vartheta_2-\vartheta_1\vartheta_2)}{1+\vartheta_1-\vartheta_0\vartheta_1+3\vartheta_2-\vartheta_0\vartheta_2-\vartheta_1\vartheta_2}\, .
\ee

We explicit the time dependence for the sake of clarity, define $\tilde{\vartheta}_j(t,y(\lambda))=\vartheta_j(t,\lambda)$ and change variables in the GHD equation using the identity $\partial_t\tilde{\vartheta}_j(t,y)=\partial_t\vartheta_j(t,\lambda)+\partial_t\lambda(y)\Big|_{y=\text{const}}\partial_\lambda \vartheta_j(t,\lambda)$.
One can read $\partial_t\lambda(y)\Big|_{y=\text{const}}$ from Eq.~\eqref{eq_ydef} and using $\partial_t\rho_0(\lambda)+\partial_\lambda (a_0^\eff\rho_0(\lambda))=0$, thus
\be\label{eq_dlambda}
\partial_t\lambda(y)\Big|_{y=\text{const}}=a_0^\eff(\lambda)\, .
\ee
In the three-string subspace, one finds $a_0^\eff(\lambda)=2\pi \hbar F(\tilde{\vartheta}_1 + 2 \tilde{\vartheta}_2 - \tilde{\vartheta}_1 \tilde{\vartheta}_2)/(1 + \tilde{\vartheta}_1 + 3 \tilde{\vartheta}_2 - \tilde{\vartheta}_1 \vartheta_2)$.
The GHD equations are transformed in $\partial_t\tilde{\vartheta}_0(y)=0$ and
\be\label{eq_bloch_weak}
\partial_t\tilde{\vartheta}_1(y)+F\frac{1-\tilde{\vartheta}_2}{1+\tilde{\vartheta}_2}\partial_y\tilde{\vartheta}_1=0\, ,\hspace{1pc}
\partial_t\tilde{\vartheta}_2(y)+F\frac{2}{1-\tilde{\vartheta}_1}\partial_y\tilde{\vartheta}_2=0\, .
\ee

If in the above equation the two-magnon string is neglected $\tilde{\vartheta}_2(y)=0$,  perfect Bloch oscillations with a constant period are obtained. For a finite value of $\tilde{\vartheta}_2(\lambda)$, the equations become coupled and non-linear and thus feature a richer frequency spectrum compared to the perfect Bloch oscillations. This can already be understood by expanding Eq. \eqref{eq_bloch_weak} for small filling $\partial_t\tilde{\vartheta}_1(y)+F(1-2\tilde{\vartheta}_2)\partial_y\tilde{\vartheta}_1\simeq0$ and then taking the average over one period of the acceleration $\partial_t\tilde{\vartheta}_1(y)+F(1-2\langle\tilde{\vartheta}_2\rangle)\partial_y\tilde{\vartheta}_1\simeq0$, with $\langle\tilde{\vartheta}_2\rangle=(2\hbar k_\text{F})^{-1}\int \dd y \tilde{\vartheta}_2(y)$ and thus expliciting a renormalization from the fundamental frequency. This spoils the perfect resonance of the oscillation frequencies of the strings $j=1$ and $j=2$ observed in the limit of zero impurity density, and gives rise to the slow beat shown in Fig.~\ref{fig_3}, where Eq.~\eqref{eq_bloch_weak} has been numerically solved with the standard method of characteristics, which we explain in details in SM~\cite{suppmat} for the numerical solution of the general GHD equation.

As the change of coordinate is comoving to the string $j=0$, the overall drift is not evident, but it can be made explicit by integrating Eq.~\eqref{eq_dlambda} in time. 
For simplicity, we consider Eq.~\eqref{eq_dlambda} neglecting two-magnon states and integrate it over one oscillation period 
\begin{multline}
\Delta \lambda\Big|_{y=\text{const}}=\int_0^T\dd t\, a_{0;t}^\eff(y)\Big|_{y=\text{const}}=\\
=2\pi\hbar\int_0^{2\hbar k_\text{F}}\dd y'\frac{\tilde{\vartheta}_1(y') }{1 + \tilde{\vartheta}_1(y') }\, .
\end{multline}
We see that after one period $\Delta \lambda\Big|_{y=\text{const}}$ is a constant shift independent from $y$, hence it describes a rigid shift of the initial occupancy in the rapidity space. Consistently with the acceleration of the center of mass, $\Delta \lambda\Big|_{y=\text{const}}=  2k_\text{F}  n_\dw/n$.

\bibliography{biblio}

\begin{thebibliography}{69}%
\makeatletter
\providecommand \@ifxundefined [1]{%
 \@ifx{#1\undefined}
}%
\providecommand \@ifnum [1]{%
 \ifnum #1\expandafter \@firstoftwo
 \else \expandafter \@secondoftwo
 \fi
}%
\providecommand \@ifx [1]{%
 \ifx #1\expandafter \@firstoftwo
 \else \expandafter \@secondoftwo
 \fi
}%
\providecommand \natexlab [1]{#1}%
\providecommand \enquote  [1]{``#1''}%
\providecommand \bibnamefont  [1]{#1}%
\providecommand \bibfnamefont [1]{#1}%
\providecommand \citenamefont [1]{#1}%
\providecommand \href@noop [0]{\@secondoftwo}%
\providecommand \href [0]{\begingroup \@sanitize@url \@href}%
\providecommand \@href[1]{\@@startlink{#1}\@@href}%
\providecommand \@@href[1]{\endgroup#1\@@endlink}%
\providecommand \@sanitize@url [0]{\catcode `\\12\catcode `\$12\catcode
  `\&12\catcode `\#12\catcode `\^12\catcode `\_12\catcode `\%12\relax}%
\providecommand \@@startlink[1]{}%
\providecommand \@@endlink[0]{}%
\providecommand \url  [0]{\begingroup\@sanitize@url \@url }%
\providecommand \@url [1]{\endgroup\@href {#1}{\urlprefix }}%
\providecommand \urlprefix  [0]{URL }%
\providecommand \Eprint [0]{\href }%
\providecommand \doibase [0]{https://doi.org/}%
\providecommand \selectlanguage [0]{\@gobble}%
\providecommand \bibinfo  [0]{\@secondoftwo}%
\providecommand \bibfield  [0]{\@secondoftwo}%
\providecommand \translation [1]{[#1]}%
\providecommand \BibitemOpen [0]{}%
\providecommand \bibitemStop [0]{}%
\providecommand \bibitemNoStop [0]{.\EOS\space}%
\providecommand \EOS [0]{\spacefactor3000\relax}%
\providecommand \BibitemShut  [1]{\csname bibitem#1\endcsname}%
\let\auto@bib@innerbib\@empty
\bibitem [{\citenamefont {Bloch}(1929)}]{bloch1929}%
  \BibitemOpen
  \bibfield  {author} {\bibinfo {author} {\bibfnamefont {F.}~\bibnamefont
  {Bloch}},\ }\bibfield  {title} {\bibinfo {title} {{\"U}ber die
  quantenmechanik der elektronen in kristallgittern},\ }\href@noop {}
  {\bibfield  {journal} {\bibinfo  {journal} {Zeitschrift f{\"u}r physik}\
  }\textbf {\bibinfo {volume} {52}},\ \bibinfo {pages} {555} (\bibinfo {year}
  {1929})}\BibitemShut {NoStop}%
\bibitem [{\citenamefont {Zener}(1934)}]{zener1934}%
  \BibitemOpen
  \bibfield  {author} {\bibinfo {author} {\bibfnamefont {C.}~\bibnamefont
  {Zener}},\ }\bibfield  {title} {\bibinfo {title} {A theory of the electrical
  breakdown of solid dielectrics},\ }\href@noop {} {\bibfield  {journal}
  {\bibinfo  {journal} {Proceedings of the Royal Society of London. Series A,
  Containing Papers of a Mathematical and Physical Character}\ }\textbf
  {\bibinfo {volume} {145}},\ \bibinfo {pages} {523} (\bibinfo {year}
  {1934})}\BibitemShut {NoStop}%
\bibitem [{\citenamefont {Ben~Dahan}\ \emph {et~al.}(1996)\citenamefont
  {Ben~Dahan}, \citenamefont {Peik}, \citenamefont {Reichel}, \citenamefont
  {Castin},\ and\ \citenamefont {Salomon}}]{Dahan1996}%
  \BibitemOpen
  \bibfield  {author} {\bibinfo {author} {\bibfnamefont {M.}~\bibnamefont
  {Ben~Dahan}}, \bibinfo {author} {\bibfnamefont {E.}~\bibnamefont {Peik}},
  \bibinfo {author} {\bibfnamefont {J.}~\bibnamefont {Reichel}}, \bibinfo
  {author} {\bibfnamefont {Y.}~\bibnamefont {Castin}},\ and\ \bibinfo {author}
  {\bibfnamefont {C.}~\bibnamefont {Salomon}},\ }\bibfield  {title} {\bibinfo
  {title} {Bloch oscillations of atoms in an optical potential},\ }\href
  {https://doi.org/10.1103/PhysRevLett.76.4508} {\bibfield  {journal} {\bibinfo
   {journal} {Phys. Rev. Lett.}\ }\textbf {\bibinfo {volume} {76}},\ \bibinfo
  {pages} {4508} (\bibinfo {year} {1996})}\BibitemShut {NoStop}%
\bibitem [{\citenamefont {Gustavsson}\ \emph {et~al.}(2008)\citenamefont
  {Gustavsson}, \citenamefont {Haller}, \citenamefont {Mark}, \citenamefont
  {Danzl}, \citenamefont {Rojas-Kopeinig},\ and\ \citenamefont
  {N\"agerl}}]{Gustavsson2008}%
  \BibitemOpen
  \bibfield  {author} {\bibinfo {author} {\bibfnamefont {M.}~\bibnamefont
  {Gustavsson}}, \bibinfo {author} {\bibfnamefont {E.}~\bibnamefont {Haller}},
  \bibinfo {author} {\bibfnamefont {M.~J.}\ \bibnamefont {Mark}}, \bibinfo
  {author} {\bibfnamefont {J.~G.}\ \bibnamefont {Danzl}}, \bibinfo {author}
  {\bibfnamefont {G.}~\bibnamefont {Rojas-Kopeinig}},\ and\ \bibinfo {author}
  {\bibfnamefont {H.-C.}\ \bibnamefont {N\"agerl}},\ }\bibfield  {title}
  {\bibinfo {title} {Control of interaction-induced dephasing of bloch
  oscillations},\ }\href {https://doi.org/10.1103/PhysRevLett.100.080404}
  {\bibfield  {journal} {\bibinfo  {journal} {Phys. Rev. Lett.}\ }\textbf
  {\bibinfo {volume} {100}},\ \bibinfo {pages} {080404} (\bibinfo {year}
  {2008})}\BibitemShut {NoStop}%
\bibitem [{\citenamefont {Schmidt}\ \emph {et~al.}(2018)\citenamefont
  {Schmidt}, \citenamefont {Knap}, \citenamefont {Ivanov}, \citenamefont {You},
  \citenamefont {Cetina},\ and\ \citenamefont {Demler}}]{Schmidt2018}%
  \BibitemOpen
  \bibfield  {author} {\bibinfo {author} {\bibfnamefont {R.}~\bibnamefont
  {Schmidt}}, \bibinfo {author} {\bibfnamefont {M.}~\bibnamefont {Knap}},
  \bibinfo {author} {\bibfnamefont {D.~A.}\ \bibnamefont {Ivanov}}, \bibinfo
  {author} {\bibfnamefont {J.-S.}\ \bibnamefont {You}}, \bibinfo {author}
  {\bibfnamefont {M.}~\bibnamefont {Cetina}},\ and\ \bibinfo {author}
  {\bibfnamefont {E.}~\bibnamefont {Demler}},\ }\bibfield  {title} {\bibinfo
  {title} {Universal many-body response of heavy impurities coupled to a fermi
  sea: a review of recent progress},\ }\href
  {https://doi.org/10.1088/1361-6633/aa9593} {\bibfield  {journal} {\bibinfo
  {journal} {Reports on Progress in Physics}\ }\textbf {\bibinfo {volume}
  {81}},\ \bibinfo {pages} {024401} (\bibinfo {year} {2018})}\BibitemShut
  {NoStop}%
\bibitem [{\citenamefont {Giamarchi}(2003)}]{giamarchi2003quantum}%
  \BibitemOpen
  \bibfield  {author} {\bibinfo {author} {\bibfnamefont {T.}~\bibnamefont
  {Giamarchi}},\ }\href@noop {} {\emph {\bibinfo {title} {Quantum physics in
  one dimension}}},\ Vol.\ \bibinfo {volume} {121}\ (\bibinfo  {publisher}
  {Clarendon press},\ \bibinfo {year} {2003})\BibitemShut {NoStop}%
\bibitem [{\citenamefont {Gangardt}\ and\ \citenamefont
  {Kamenev}(2009)}]{Gangardt2009}%
  \BibitemOpen
  \bibfield  {author} {\bibinfo {author} {\bibfnamefont {D.~M.}\ \bibnamefont
  {Gangardt}}\ and\ \bibinfo {author} {\bibfnamefont {A.}~\bibnamefont
  {Kamenev}},\ }\bibfield  {title} {\bibinfo {title} {Bloch oscillations in a
  one-dimensional spinor gas},\ }\href
  {https://doi.org/10.1103/PhysRevLett.102.070402} {\bibfield  {journal}
  {\bibinfo  {journal} {Phys. Rev. Lett.}\ }\textbf {\bibinfo {volume} {102}},\
  \bibinfo {pages} {070402} (\bibinfo {year} {2009})}\BibitemShut {NoStop}%
\bibitem [{\citenamefont {Matveev}\ and\ \citenamefont
  {Furusaki}(2008)}]{Matveev2008}%
  \BibitemOpen
  \bibfield  {author} {\bibinfo {author} {\bibfnamefont {K.~A.}\ \bibnamefont
  {Matveev}}\ and\ \bibinfo {author} {\bibfnamefont {A.}~\bibnamefont
  {Furusaki}},\ }\bibfield  {title} {\bibinfo {title} {Spectral functions of
  strongly interacting isospin-$\frac{1}{2}$ bosons in one dimension},\ }\href
  {https://doi.org/10.1103/PhysRevLett.101.170403} {\bibfield  {journal}
  {\bibinfo  {journal} {Phys. Rev. Lett.}\ }\textbf {\bibinfo {volume} {101}},\
  \bibinfo {pages} {170403} (\bibinfo {year} {2008})}\BibitemShut {NoStop}%
\bibitem [{\citenamefont {Kamenev}\ and\ \citenamefont
  {Glazman}(2009)}]{Kamenev2009}%
  \BibitemOpen
  \bibfield  {author} {\bibinfo {author} {\bibfnamefont {A.}~\bibnamefont
  {Kamenev}}\ and\ \bibinfo {author} {\bibfnamefont {L.~I.}\ \bibnamefont
  {Glazman}},\ }\bibfield  {title} {\bibinfo {title} {Dynamics of a
  one-dimensional spinor bose liquid: A phenomenological approach},\ }\href
  {https://doi.org/10.1103/PhysRevA.80.011603} {\bibfield  {journal} {\bibinfo
  {journal} {Phys. Rev. A}\ }\textbf {\bibinfo {volume} {80}},\ \bibinfo
  {pages} {011603} (\bibinfo {year} {2009})}\BibitemShut {NoStop}%
\bibitem [{\citenamefont {Schecter}\ \emph {et~al.}(2012)\citenamefont
  {Schecter}, \citenamefont {Gangardt},\ and\ \citenamefont
  {Kamenev}}]{Schecter2012}%
  \BibitemOpen
  \bibfield  {author} {\bibinfo {author} {\bibfnamefont {M.}~\bibnamefont
  {Schecter}}, \bibinfo {author} {\bibfnamefont {D.}~\bibnamefont {Gangardt}},\
  and\ \bibinfo {author} {\bibfnamefont {A.}~\bibnamefont {Kamenev}},\
  }\bibfield  {title} {\bibinfo {title} {Dynamics and bloch oscillations of
  mobile impurities in one-dimensional quantum liquids},\ }\href
  {https://doi.org/https://doi.org/10.1016/j.aop.2011.10.001} {\bibfield
  {journal} {\bibinfo  {journal} {Annals of Physics}\ }\textbf {\bibinfo
  {volume} {327}},\ \bibinfo {pages} {639} (\bibinfo {year}
  {2012})}\BibitemShut {NoStop}%
\bibitem [{\citenamefont {Gamayun}\ \emph {et~al.}(2014)\citenamefont
  {Gamayun}, \citenamefont {Lychkovskiy},\ and\ \citenamefont
  {Cheianov}}]{Gamayun2014}%
  \BibitemOpen
  \bibfield  {author} {\bibinfo {author} {\bibfnamefont {O.}~\bibnamefont
  {Gamayun}}, \bibinfo {author} {\bibfnamefont {O.}~\bibnamefont
  {Lychkovskiy}},\ and\ \bibinfo {author} {\bibfnamefont {V.}~\bibnamefont
  {Cheianov}},\ }\bibfield  {title} {\bibinfo {title} {Kinetic theory for a
  mobile impurity in a degenerate tonks-girardeau gas},\ }\href
  {https://doi.org/10.1103/PhysRevE.90.032132} {\bibfield  {journal} {\bibinfo
  {journal} {Phys. Rev. E}\ }\textbf {\bibinfo {volume} {90}},\ \bibinfo
  {pages} {032132} (\bibinfo {year} {2014})}\BibitemShut {NoStop}%
\bibitem [{\citenamefont {Schecter}\ \emph {et~al.}(2015)\citenamefont
  {Schecter}, \citenamefont {Gangardt},\ and\ \citenamefont
  {Kamenev}}]{Schecter2015}%
  \BibitemOpen
  \bibfield  {author} {\bibinfo {author} {\bibfnamefont {M.}~\bibnamefont
  {Schecter}}, \bibinfo {author} {\bibfnamefont {D.~M.}\ \bibnamefont
  {Gangardt}},\ and\ \bibinfo {author} {\bibfnamefont {A.}~\bibnamefont
  {Kamenev}},\ }\bibfield  {title} {\bibinfo {title} {Comment on ``kinetic
  theory for a mobile impurity in a degenerate tonks-girardeau gas''},\ }\href
  {https://doi.org/10.1103/PhysRevE.92.016101} {\bibfield  {journal} {\bibinfo
  {journal} {Phys. Rev. E}\ }\textbf {\bibinfo {volume} {92}},\ \bibinfo
  {pages} {016101} (\bibinfo {year} {2015})}\BibitemShut {NoStop}%
\bibitem [{\citenamefont {Gamayun}\ \emph {et~al.}(2015)\citenamefont
  {Gamayun}, \citenamefont {Lychkovskiy},\ and\ \citenamefont
  {Cheianov}}]{Gamayun2015}%
  \BibitemOpen
  \bibfield  {author} {\bibinfo {author} {\bibfnamefont {O.}~\bibnamefont
  {Gamayun}}, \bibinfo {author} {\bibfnamefont {O.}~\bibnamefont
  {Lychkovskiy}},\ and\ \bibinfo {author} {\bibfnamefont {V.}~\bibnamefont
  {Cheianov}},\ }\bibfield  {title} {\bibinfo {title} {Reply to ``comment on
  `kinetic theory for a mobile impurity in a degenerate tonks-girardeau gas'
  ''},\ }\href {https://doi.org/10.1103/PhysRevE.92.016102} {\bibfield
  {journal} {\bibinfo  {journal} {Phys. Rev. E}\ }\textbf {\bibinfo {volume}
  {92}},\ \bibinfo {pages} {016102} (\bibinfo {year} {2015})}\BibitemShut
  {NoStop}%
\bibitem [{\citenamefont {Yang}\ \emph {et~al.}(2017)\citenamefont {Yang},
  \citenamefont {Zhou}, \citenamefont {Yi},\ and\ \citenamefont
  {Cui}}]{Yang2017}%
  \BibitemOpen
  \bibfield  {author} {\bibinfo {author} {\bibfnamefont {L.}~\bibnamefont
  {Yang}}, \bibinfo {author} {\bibfnamefont {L.}~\bibnamefont {Zhou}}, \bibinfo
  {author} {\bibfnamefont {W.}~\bibnamefont {Yi}},\ and\ \bibinfo {author}
  {\bibfnamefont {X.}~\bibnamefont {Cui}},\ }\bibfield  {title} {\bibinfo
  {title} {Interaction-induced bloch oscillation in a harmonically trapped and
  fermionized quantum gas in one dimension},\ }\href
  {https://doi.org/10.1103/PhysRevA.95.053617} {\bibfield  {journal} {\bibinfo
  {journal} {Phys. Rev. A}\ }\textbf {\bibinfo {volume} {95}},\ \bibinfo
  {pages} {053617} (\bibinfo {year} {2017})}\BibitemShut {NoStop}%
\bibitem [{\citenamefont {Lychkovskiy}\ \emph {et~al.}(2018)\citenamefont
  {Lychkovskiy}, \citenamefont {Gamayun},\ and\ \citenamefont
  {Cheianov}}]{Lychkovskiy2018}%
  \BibitemOpen
  \bibfield  {author} {\bibinfo {author} {\bibfnamefont {O.}~\bibnamefont
  {Lychkovskiy}}, \bibinfo {author} {\bibfnamefont {O.}~\bibnamefont
  {Gamayun}},\ and\ \bibinfo {author} {\bibfnamefont {V.}~\bibnamefont
  {Cheianov}},\ }\bibfield  {title} {\bibinfo {title} {Necessary and sufficient
  condition for quantum adiabaticity in a driven one-dimensional impurity-fluid
  system},\ }\href {https://doi.org/10.1103/PhysRevB.98.024307} {\bibfield
  {journal} {\bibinfo  {journal} {Phys. Rev. B}\ }\textbf {\bibinfo {volume}
  {98}},\ \bibinfo {pages} {024307} (\bibinfo {year} {2018})}\BibitemShut
  {NoStop}%
\bibitem [{\citenamefont {Majumdar}\ and\ \citenamefont
  {Petkovi\ifmmode~\acute{c}\else \'{c}\fi{}}(2026)}]{Majumdar2026}%
  \BibitemOpen
  \bibfield  {author} {\bibinfo {author} {\bibfnamefont {S.}~\bibnamefont
  {Majumdar}}\ and\ \bibinfo {author} {\bibfnamefont {A.}~\bibnamefont
  {Petkovi\ifmmode~\acute{c}\else \'{c}\fi{}}},\ }\bibfield  {title} {\bibinfo
  {title} {Bloch oscillations of a mobile impurity in a one-dimensional bose
  gas},\ }\href {https://doi.org/10.1103/bfcg-m2tw} {\bibfield  {journal}
  {\bibinfo  {journal} {Phys. Rev. A}\ }\textbf {\bibinfo {volume} {113}},\
  \bibinfo {pages} {043309} (\bibinfo {year} {2026})}\BibitemShut {NoStop}%
\bibitem [{\citenamefont {Meinert}\ \emph {et~al.}(2017)\citenamefont
  {Meinert}, \citenamefont {Knap}, \citenamefont {Kirilov}, \citenamefont
  {Jag-Lauber}, \citenamefont {Zvonarev}, \citenamefont {Demler},\ and\
  \citenamefont {Nägerl}}]{Meinert2017}%
  \BibitemOpen
  \bibfield  {author} {\bibinfo {author} {\bibfnamefont {F.}~\bibnamefont
  {Meinert}}, \bibinfo {author} {\bibfnamefont {M.}~\bibnamefont {Knap}},
  \bibinfo {author} {\bibfnamefont {E.}~\bibnamefont {Kirilov}}, \bibinfo
  {author} {\bibfnamefont {K.}~\bibnamefont {Jag-Lauber}}, \bibinfo {author}
  {\bibfnamefont {M.~B.}\ \bibnamefont {Zvonarev}}, \bibinfo {author}
  {\bibfnamefont {E.}~\bibnamefont {Demler}},\ and\ \bibinfo {author}
  {\bibfnamefont {H.-C.}\ \bibnamefont {Nägerl}},\ }\bibfield  {title}
  {\bibinfo {title} {Bloch oscillations in the absence of a lattice},\ }\href
  {https://doi.org/10.1126/science.aah6616} {\bibfield  {journal} {\bibinfo
  {journal} {Science}\ }\textbf {\bibinfo {volume} {356}},\ \bibinfo {pages}
  {945} (\bibinfo {year} {2017})},\ \Eprint
  {https://arxiv.org/abs/https://www.science.org/doi/pdf/10.1126/science.aah6616}
  {https://www.science.org/doi/pdf/10.1126/science.aah6616} \BibitemShut
  {NoStop}%
\bibitem [{\citenamefont {Rabec}\ \emph {et~al.}(2025)\citenamefont {Rabec},
  \citenamefont {Chauveau}, \citenamefont {Brochier}, \citenamefont
  {Nascimbene}, \citenamefont {Dalibard},\ and\ \citenamefont
  {Beugnon}}]{Rabec2025}%
  \BibitemOpen
  \bibfield  {author} {\bibinfo {author} {\bibfnamefont {F.}~\bibnamefont
  {Rabec}}, \bibinfo {author} {\bibfnamefont {G.}~\bibnamefont {Chauveau}},
  \bibinfo {author} {\bibfnamefont {G.}~\bibnamefont {Brochier}}, \bibinfo
  {author} {\bibfnamefont {S.}~\bibnamefont {Nascimbene}}, \bibinfo {author}
  {\bibfnamefont {J.}~\bibnamefont {Dalibard}},\ and\ \bibinfo {author}
  {\bibfnamefont {J.}~\bibnamefont {Beugnon}},\ }\bibfield  {title} {\bibinfo
  {title} {Bloch oscillations of a soliton in a one-dimensional quantum
  fluid},\ }\href {https://doi.org/10.1038/s41567-025-02970-1} {\bibfield
  {journal} {\bibinfo  {journal} {Nature Physics}\ }\textbf {\bibinfo {volume}
  {21}},\ \bibinfo {pages} {1541} (\bibinfo {year} {2025})}\BibitemShut
  {NoStop}%
\bibitem [{\citenamefont {Berry}(2009)}]{Berry2009}%
  \BibitemOpen
  \bibfield  {author} {\bibinfo {author} {\bibfnamefont {M.~V.}\ \bibnamefont
  {Berry}},\ }\bibfield  {title} {\bibinfo {title} {Transitionless quantum
  driving},\ }\href {https://doi.org/10.1088/1751-8113/42/36/365303} {\bibfield
   {journal} {\bibinfo  {journal} {Journal of Physics A: Mathematical and
  Theoretical}\ }\textbf {\bibinfo {volume} {42}},\ \bibinfo {pages} {365303}
  (\bibinfo {year} {2009})}\BibitemShut {NoStop}%
\bibitem [{\citenamefont {Takahashi}(2005)}]{takahashi2005thermodynamics}%
  \BibitemOpen
  \bibfield  {author} {\bibinfo {author} {\bibfnamefont {M.}~\bibnamefont
  {Takahashi}},\ }\href@noop {} {\emph {\bibinfo {title} {Thermodynamics of
  one-dimensional solvable models}}}\ (\bibinfo  {publisher} {Cambridge
  University Press},\ \bibinfo {year} {2005})\BibitemShut {NoStop}%
\bibitem [{\citenamefont {Rigol}\ \emph {et~al.}(2007)\citenamefont {Rigol},
  \citenamefont {Dunjko}, \citenamefont {Yurovsky},\ and\ \citenamefont
  {Olshanii}}]{Rigol2009}%
  \BibitemOpen
  \bibfield  {author} {\bibinfo {author} {\bibfnamefont {M.}~\bibnamefont
  {Rigol}}, \bibinfo {author} {\bibfnamefont {V.}~\bibnamefont {Dunjko}},
  \bibinfo {author} {\bibfnamefont {V.}~\bibnamefont {Yurovsky}},\ and\
  \bibinfo {author} {\bibfnamefont {M.}~\bibnamefont {Olshanii}},\ }\bibfield
  {title} {\bibinfo {title} {Relaxation in a completely integrable many-body
  quantum system: An ab initio study of the dynamics of the highly excited
  states of 1d lattice hard-core bosons},\ }\href
  {https://doi.org/10.1103/PhysRevLett.98.050405} {\bibfield  {journal}
  {\bibinfo  {journal} {Phys. Rev. Lett.}\ }\textbf {\bibinfo {volume} {98}},\
  \bibinfo {pages} {050405} (\bibinfo {year} {2007})}\BibitemShut {NoStop}%
\bibitem [{\citenamefont {Langen}\ \emph {et~al.}(2015)\citenamefont {Langen},
  \citenamefont {Erne}, \citenamefont {Geiger}, \citenamefont {Rauer},
  \citenamefont {Schweigler}, \citenamefont {Kuhnert}, \citenamefont
  {Rohringer}, \citenamefont {Mazets}, \citenamefont {Gasenzer},\ and\
  \citenamefont {Schmiedmayer}}]{Langen2015}%
  \BibitemOpen
  \bibfield  {author} {\bibinfo {author} {\bibfnamefont {T.}~\bibnamefont
  {Langen}}, \bibinfo {author} {\bibfnamefont {S.}~\bibnamefont {Erne}},
  \bibinfo {author} {\bibfnamefont {R.}~\bibnamefont {Geiger}}, \bibinfo
  {author} {\bibfnamefont {B.}~\bibnamefont {Rauer}}, \bibinfo {author}
  {\bibfnamefont {T.}~\bibnamefont {Schweigler}}, \bibinfo {author}
  {\bibfnamefont {M.}~\bibnamefont {Kuhnert}}, \bibinfo {author} {\bibfnamefont
  {W.}~\bibnamefont {Rohringer}}, \bibinfo {author} {\bibfnamefont {I.~E.}\
  \bibnamefont {Mazets}}, \bibinfo {author} {\bibfnamefont {T.}~\bibnamefont
  {Gasenzer}},\ and\ \bibinfo {author} {\bibfnamefont {J.}~\bibnamefont
  {Schmiedmayer}},\ }\bibfield  {title} {\bibinfo {title} {Experimental
  observation of a generalized gibbs ensemble},\ }\href
  {https://doi.org/10.1126/science.1257026} {\bibfield  {journal} {\bibinfo
  {journal} {Science}\ }\textbf {\bibinfo {volume} {348}},\ \bibinfo {pages}
  {207} (\bibinfo {year} {2015})},\ \Eprint
  {https://arxiv.org/abs/https://www.science.org/doi/pdf/10.1126/science.1257026}
  {https://www.science.org/doi/pdf/10.1126/science.1257026} \BibitemShut
  {NoStop}%
\bibitem [{\citenamefont {Calabrese}\ \emph {et~al.}(2016)\citenamefont
  {Calabrese}, \citenamefont {Essler},\ and\ \citenamefont
  {Mussardo}}]{Calabrese2016}%
  \BibitemOpen
  \bibfield  {author} {\bibinfo {author} {\bibfnamefont {P.}~\bibnamefont
  {Calabrese}}, \bibinfo {author} {\bibfnamefont {F.~H.~L.}\ \bibnamefont
  {Essler}},\ and\ \bibinfo {author} {\bibfnamefont {G.}~\bibnamefont
  {Mussardo}},\ }\bibfield  {title} {\bibinfo {title} {Introduction to
  ‘quantum integrability in out of equilibrium systems’},\ }\href
  {https://doi.org/10.1088/1742-5468/2016/06/064001} {\bibfield  {journal}
  {\bibinfo  {journal} {Journal of Statistical Mechanics: Theory and
  Experiment}\ }\textbf {\bibinfo {volume} {2016}},\ \bibinfo {pages} {064001}
  (\bibinfo {year} {2016})}\BibitemShut {NoStop}%
\bibitem [{\citenamefont {Guan}\ and\ \citenamefont {He}(2022)}]{Guan2022}%
  \BibitemOpen
  \bibfield  {author} {\bibinfo {author} {\bibfnamefont {X.-W.}\ \bibnamefont
  {Guan}}\ and\ \bibinfo {author} {\bibfnamefont {P.}~\bibnamefont {He}},\
  }\bibfield  {title} {\bibinfo {title} {New trends in quantum integrability:
  recent experiments with ultracold atoms},\ }\href
  {https://doi.org/10.1088/1361-6633/ac95a9} {\bibfield  {journal} {\bibinfo
  {journal} {Reports on Progress in Physics}\ }\textbf {\bibinfo {volume}
  {85}},\ \bibinfo {pages} {114001} (\bibinfo {year} {2022})}\BibitemShut
  {NoStop}%
\bibitem [{\citenamefont {Castro-Alvaredo}\ \emph {et~al.}(2016)\citenamefont
  {Castro-Alvaredo}, \citenamefont {Doyon},\ and\ \citenamefont
  {Yoshimura}}]{Alvaredo2016}%
  \BibitemOpen
  \bibfield  {author} {\bibinfo {author} {\bibfnamefont {O.~A.}\ \bibnamefont
  {Castro-Alvaredo}}, \bibinfo {author} {\bibfnamefont {B.}~\bibnamefont
  {Doyon}},\ and\ \bibinfo {author} {\bibfnamefont {T.}~\bibnamefont
  {Yoshimura}},\ }\bibfield  {title} {\bibinfo {title} {Emergent hydrodynamics
  in integrable quantum systems out of equilibrium},\ }\href
  {https://doi.org/10.1103/PhysRevX.6.041065} {\bibfield  {journal} {\bibinfo
  {journal} {Phys. Rev. X}\ }\textbf {\bibinfo {volume} {6}},\ \bibinfo {pages}
  {041065} (\bibinfo {year} {2016})}\BibitemShut {NoStop}%
\bibitem [{\citenamefont {Bertini}\ \emph {et~al.}(2016)\citenamefont
  {Bertini}, \citenamefont {Collura}, \citenamefont {De~Nardis},\ and\
  \citenamefont {Fagotti}}]{Bertini2016}%
  \BibitemOpen
  \bibfield  {author} {\bibinfo {author} {\bibfnamefont {B.}~\bibnamefont
  {Bertini}}, \bibinfo {author} {\bibfnamefont {M.}~\bibnamefont {Collura}},
  \bibinfo {author} {\bibfnamefont {J.}~\bibnamefont {De~Nardis}},\ and\
  \bibinfo {author} {\bibfnamefont {M.}~\bibnamefont {Fagotti}},\ }\bibfield
  {title} {\bibinfo {title} {Transport in out-of-equilibrium xxz chains: Exact
  profiles of charges and currents},\ }\href
  {https://doi.org/10.1103/PhysRevLett.117.207201} {\bibfield  {journal}
  {\bibinfo  {journal} {Phys. Rev. Lett.}\ }\textbf {\bibinfo {volume} {117}},\
  \bibinfo {pages} {207201} (\bibinfo {year} {2016})}\BibitemShut {NoStop}%
\bibitem [{\citenamefont {Bastianello}\ \emph {et~al.}(2022)\citenamefont
  {Bastianello}, \citenamefont {Bertini}, \citenamefont {Doyon},\ and\
  \citenamefont {Vasseur}}]{Bastianello2022}%
  \BibitemOpen
  \bibfield  {author} {\bibinfo {author} {\bibfnamefont {A.}~\bibnamefont
  {Bastianello}}, \bibinfo {author} {\bibfnamefont {B.}~\bibnamefont
  {Bertini}}, \bibinfo {author} {\bibfnamefont {B.}~\bibnamefont {Doyon}},\
  and\ \bibinfo {author} {\bibfnamefont {R.}~\bibnamefont {Vasseur}},\
  }\bibfield  {title} {\bibinfo {title} {Introduction to the special issue on
  emergent hydrodynamics in integrable many-body systems},\ }\href
  {https://doi.org/10.1088/1742-5468/ac3e6a} {\bibfield  {journal} {\bibinfo
  {journal} {Journal of Statistical Mechanics: Theory and Experiment}\ }\textbf
  {\bibinfo {volume} {2022}},\ \bibinfo {pages} {014001} (\bibinfo {year}
  {2022})}\BibitemShut {NoStop}%
\bibitem [{\citenamefont {Doyon}\ \emph {et~al.}(2025)\citenamefont {Doyon},
  \citenamefont {Gopalakrishnan}, \citenamefont {M\o{}ller}, \citenamefont
  {Schmiedmayer},\ and\ \citenamefont {Vasseur}}]{Doyon2024}%
  \BibitemOpen
  \bibfield  {author} {\bibinfo {author} {\bibfnamefont {B.}~\bibnamefont
  {Doyon}}, \bibinfo {author} {\bibfnamefont {S.}~\bibnamefont
  {Gopalakrishnan}}, \bibinfo {author} {\bibfnamefont {F.}~\bibnamefont
  {M\o{}ller}}, \bibinfo {author} {\bibfnamefont {J.}~\bibnamefont
  {Schmiedmayer}},\ and\ \bibinfo {author} {\bibfnamefont {R.}~\bibnamefont
  {Vasseur}},\ }\bibfield  {title} {\bibinfo {title} {Generalized
  hydrodynamics: a perspective},\ }\href
  {https://doi.org/10.1103/PhysRevX.15.010501} {\bibfield  {journal} {\bibinfo
  {journal} {Phys. Rev. X}\ }\textbf {\bibinfo {volume} {15}},\ \bibinfo
  {pages} {010501} (\bibinfo {year} {2025})}\BibitemShut {NoStop}%
\bibitem [{\citenamefont {Schemmer}\ \emph {et~al.}(2019)\citenamefont
  {Schemmer}, \citenamefont {Bouchoule}, \citenamefont {Doyon},\ and\
  \citenamefont {Dubail}}]{Schemmer2019}%
  \BibitemOpen
  \bibfield  {author} {\bibinfo {author} {\bibfnamefont {M.}~\bibnamefont
  {Schemmer}}, \bibinfo {author} {\bibfnamefont {I.}~\bibnamefont {Bouchoule}},
  \bibinfo {author} {\bibfnamefont {B.}~\bibnamefont {Doyon}},\ and\ \bibinfo
  {author} {\bibfnamefont {J.}~\bibnamefont {Dubail}},\ }\bibfield  {title}
  {\bibinfo {title} {Generalized hydrodynamics on an atom chip},\ }\href
  {https://doi.org/10.1103/PhysRevLett.122.090601} {\bibfield  {journal}
  {\bibinfo  {journal} {Phys. Rev. Lett.}\ }\textbf {\bibinfo {volume} {122}},\
  \bibinfo {pages} {090601} (\bibinfo {year} {2019})}\BibitemShut {NoStop}%
\bibitem [{\citenamefont {Malvania}\ \emph {et~al.}(2021)\citenamefont
  {Malvania}, \citenamefont {Zhang}, \citenamefont {Le}, \citenamefont
  {Dubail}, \citenamefont {Rigol},\ and\ \citenamefont {Weiss}}]{Malvania2021}%
  \BibitemOpen
  \bibfield  {author} {\bibinfo {author} {\bibfnamefont {N.}~\bibnamefont
  {Malvania}}, \bibinfo {author} {\bibfnamefont {Y.}~\bibnamefont {Zhang}},
  \bibinfo {author} {\bibfnamefont {Y.}~\bibnamefont {Le}}, \bibinfo {author}
  {\bibfnamefont {J.}~\bibnamefont {Dubail}}, \bibinfo {author} {\bibfnamefont
  {M.}~\bibnamefont {Rigol}},\ and\ \bibinfo {author} {\bibfnamefont {D.~S.}\
  \bibnamefont {Weiss}},\ }\bibfield  {title} {\bibinfo {title} {{Generalized
  hydrodynamics in strongly interacting 1D {B}ose gases}},\ }\href
  {https://doi.org/10.1126/science.abf0147} {\bibfield  {journal} {\bibinfo
  {journal} {Science}\ }\textbf {\bibinfo {volume} {373}},\ \bibinfo {pages}
  {1129} (\bibinfo {year} {2021})}\BibitemShut {NoStop}%
\bibitem [{\citenamefont {Schüttelkopf}\ \emph {et~al.}(2026)\citenamefont
  {Schüttelkopf}, \citenamefont {Tajik}, \citenamefont {Bazhan}, \citenamefont
  {Cataldini}, \citenamefont {Ji}, \citenamefont {Schmiedmayer},\ and\
  \citenamefont {Møller}}]{Schuttelkopf2024}%
  \BibitemOpen
  \bibfield  {author} {\bibinfo {author} {\bibfnamefont {P.}~\bibnamefont
  {Schüttelkopf}}, \bibinfo {author} {\bibfnamefont {M.}~\bibnamefont
  {Tajik}}, \bibinfo {author} {\bibfnamefont {N.}~\bibnamefont {Bazhan}},
  \bibinfo {author} {\bibfnamefont {F.}~\bibnamefont {Cataldini}}, \bibinfo
  {author} {\bibfnamefont {S.-C.}\ \bibnamefont {Ji}}, \bibinfo {author}
  {\bibfnamefont {J.}~\bibnamefont {Schmiedmayer}},\ and\ \bibinfo {author}
  {\bibfnamefont {F.}~\bibnamefont {Møller}},\ }\bibfield  {title} {\bibinfo
  {title} {Characterizing transport in a quantum gas by measuring {D}rude
  weights},\ }\href {https://doi.org/10.1126/science.ads8327} {\bibfield
  {journal} {\bibinfo  {journal} {Science}\ }\textbf {\bibinfo {volume}
  {391}},\ \bibinfo {pages} {290} (\bibinfo {year} {2026})}\BibitemShut
  {NoStop}%
\bibitem [{\citenamefont {Dubois}\ \emph {et~al.}(2024)\citenamefont {Dubois},
  \citenamefont {Th\'em\`eze}, \citenamefont {Nogrette}, \citenamefont
  {Dubail},\ and\ \citenamefont {Bouchoule}}]{dubois2024}%
  \BibitemOpen
  \bibfield  {author} {\bibinfo {author} {\bibfnamefont {L.}~\bibnamefont
  {Dubois}}, \bibinfo {author} {\bibfnamefont {G.}~\bibnamefont {Th\'em\`eze}},
  \bibinfo {author} {\bibfnamefont {F.}~\bibnamefont {Nogrette}}, \bibinfo
  {author} {\bibfnamefont {J.}~\bibnamefont {Dubail}},\ and\ \bibinfo {author}
  {\bibfnamefont {I.}~\bibnamefont {Bouchoule}},\ }\bibfield  {title} {\bibinfo
  {title} {Probing the local rapidity distribution of a one-dimensional {B}ose
  gas},\ }\href {https://doi.org/10.1103/PhysRevLett.133.113402} {\bibfield
  {journal} {\bibinfo  {journal} {Phys. Rev. Lett.}\ }\textbf {\bibinfo
  {volume} {133}},\ \bibinfo {pages} {113402} (\bibinfo {year}
  {2024})}\BibitemShut {NoStop}%
\bibitem [{\citenamefont {Cataldini}\ \emph {et~al.}(2022)\citenamefont
  {Cataldini}, \citenamefont {M\o{}ller}, \citenamefont {Tajik}, \citenamefont
  {Sabino}, \citenamefont {Ji}, \citenamefont {Mazets}, \citenamefont
  {Schweigler}, \citenamefont {Rauer},\ and\ \citenamefont
  {Schmiedmayer}}]{Cataldini2022}%
  \BibitemOpen
  \bibfield  {author} {\bibinfo {author} {\bibfnamefont {F.}~\bibnamefont
  {Cataldini}}, \bibinfo {author} {\bibfnamefont {F.}~\bibnamefont
  {M\o{}ller}}, \bibinfo {author} {\bibfnamefont {M.}~\bibnamefont {Tajik}},
  \bibinfo {author} {\bibfnamefont {J.~a.}\ \bibnamefont {Sabino}}, \bibinfo
  {author} {\bibfnamefont {S.-C.}\ \bibnamefont {Ji}}, \bibinfo {author}
  {\bibfnamefont {I.}~\bibnamefont {Mazets}}, \bibinfo {author} {\bibfnamefont
  {T.}~\bibnamefont {Schweigler}}, \bibinfo {author} {\bibfnamefont
  {B.}~\bibnamefont {Rauer}},\ and\ \bibinfo {author} {\bibfnamefont
  {J.}~\bibnamefont {Schmiedmayer}},\ }\bibfield  {title} {\bibinfo {title}
  {Emergent {P}auli blocking in a weakly interacting {B}ose gas},\ }\href
  {https://doi.org/10.1103/PhysRevX.12.041032} {\bibfield  {journal} {\bibinfo
  {journal} {Phys. Rev. X}\ }\textbf {\bibinfo {volume} {12}},\ \bibinfo
  {pages} {041032} (\bibinfo {year} {2022})}\BibitemShut {NoStop}%
\bibitem [{\citenamefont {M\o{}ller}\ \emph {et~al.}(2021)\citenamefont
  {M\o{}ller}, \citenamefont {Li}, \citenamefont {Mazets}, \citenamefont
  {Stimming}, \citenamefont {Zhou}, \citenamefont {Zhu}, \citenamefont {Chen},\
  and\ \citenamefont {Schmiedmayer}}]{Moller2021}%
  \BibitemOpen
  \bibfield  {author} {\bibinfo {author} {\bibfnamefont {F.}~\bibnamefont
  {M\o{}ller}}, \bibinfo {author} {\bibfnamefont {C.}~\bibnamefont {Li}},
  \bibinfo {author} {\bibfnamefont {I.}~\bibnamefont {Mazets}}, \bibinfo
  {author} {\bibfnamefont {H.-P.}\ \bibnamefont {Stimming}}, \bibinfo {author}
  {\bibfnamefont {T.}~\bibnamefont {Zhou}}, \bibinfo {author} {\bibfnamefont
  {Z.}~\bibnamefont {Zhu}}, \bibinfo {author} {\bibfnamefont {X.}~\bibnamefont
  {Chen}},\ and\ \bibinfo {author} {\bibfnamefont {J.}~\bibnamefont
  {Schmiedmayer}},\ }\bibfield  {title} {\bibinfo {title} {Extension of the
  generalized hydrodynamics to the dimensional crossover regime},\ }\href
  {https://doi.org/10.1103/PhysRevLett.126.090602} {\bibfield  {journal}
  {\bibinfo  {journal} {Phys. Rev. Lett.}\ }\textbf {\bibinfo {volume} {126}},\
  \bibinfo {pages} {090602} (\bibinfo {year} {2021})}\BibitemShut {NoStop}%
\bibitem [{\citenamefont {Yang}\ \emph {et~al.}(2024)\citenamefont {Yang},
  \citenamefont {Zhang}, \citenamefont {Li}, \citenamefont {Lin}, \citenamefont
  {Gopalakrishnan}, \citenamefont {Rigol},\ and\ \citenamefont
  {Lev}}]{Yang2024Phantom}%
  \BibitemOpen
  \bibfield  {author} {\bibinfo {author} {\bibfnamefont {K.}~\bibnamefont
  {Yang}}, \bibinfo {author} {\bibfnamefont {Y.}~\bibnamefont {Zhang}},
  \bibinfo {author} {\bibfnamefont {K.-Y.}\ \bibnamefont {Li}}, \bibinfo
  {author} {\bibfnamefont {K.-Y.}\ \bibnamefont {Lin}}, \bibinfo {author}
  {\bibfnamefont {S.}~\bibnamefont {Gopalakrishnan}}, \bibinfo {author}
  {\bibfnamefont {M.}~\bibnamefont {Rigol}},\ and\ \bibinfo {author}
  {\bibfnamefont {B.~L.}\ \bibnamefont {Lev}},\ }\bibfield  {title} {\bibinfo
  {title} {Phantom energy in the nonlinear response of a quantum many-body scar
  state},\ }\href {https://doi.org/10.1126/science.adk8978} {\bibfield
  {journal} {\bibinfo  {journal} {Science}\ }\textbf {\bibinfo {volume}
  {385}},\ \bibinfo {pages} {1063} (\bibinfo {year} {2024})}\BibitemShut
  {NoStop}%
\bibitem [{\citenamefont {Horvath}\ \emph {et~al.}(2025)\citenamefont
  {Horvath}, \citenamefont {Bastianello}, \citenamefont {Dhar}, \citenamefont
  {Koch}, \citenamefont {Guo}, \citenamefont {Caux}, \citenamefont {Landini},\
  and\ \citenamefont {Nägerl}}]{horvath2025}%
  \BibitemOpen
  \bibfield  {author} {\bibinfo {author} {\bibfnamefont {M.}~\bibnamefont
  {Horvath}}, \bibinfo {author} {\bibfnamefont {A.}~\bibnamefont
  {Bastianello}}, \bibinfo {author} {\bibfnamefont {S.}~\bibnamefont {Dhar}},
  \bibinfo {author} {\bibfnamefont {R.}~\bibnamefont {Koch}}, \bibinfo {author}
  {\bibfnamefont {Y.}~\bibnamefont {Guo}}, \bibinfo {author} {\bibfnamefont
  {J.-S.}\ \bibnamefont {Caux}}, \bibinfo {author} {\bibfnamefont
  {M.}~\bibnamefont {Landini}},\ and\ \bibinfo {author} {\bibfnamefont {H.-C.}\
  \bibnamefont {Nägerl}},\ }\href {https://arxiv.org/abs/2505.10550} {\bibinfo
  {title} {Observing bethe strings in an attractive {B}ose gas far from
  equilibrium}} (\bibinfo {year} {2025}),\ \Eprint
  {https://arxiv.org/abs/2505.10550} {arXiv:2505.10550} \BibitemShut {NoStop}%
\bibitem [{\citenamefont {Zeng}\ \emph {et~al.}(2026)\citenamefont {Zeng},
  \citenamefont {Bastianello}, \citenamefont {Dhar}, \citenamefont {Wang},
  \citenamefont {Yu}, \citenamefont {Horvath}, \citenamefont {Astrakharchik},
  \citenamefont {Guo}, \citenamefont {Nägerl},\ and\ \citenamefont
  {Landini}}]{Zeng2026}%
  \BibitemOpen
  \bibfield  {author} {\bibinfo {author} {\bibfnamefont {Y.}~\bibnamefont
  {Zeng}}, \bibinfo {author} {\bibfnamefont {A.}~\bibnamefont {Bastianello}},
  \bibinfo {author} {\bibfnamefont {S.}~\bibnamefont {Dhar}}, \bibinfo {author}
  {\bibfnamefont {Z.}~\bibnamefont {Wang}}, \bibinfo {author} {\bibfnamefont
  {X.}~\bibnamefont {Yu}}, \bibinfo {author} {\bibfnamefont {M.}~\bibnamefont
  {Horvath}}, \bibinfo {author} {\bibfnamefont {G.~E.}\ \bibnamefont
  {Astrakharchik}}, \bibinfo {author} {\bibfnamefont {Y.}~\bibnamefont {Guo}},
  \bibinfo {author} {\bibfnamefont {H.-C.}\ \bibnamefont {Nägerl}},\ and\
  \bibinfo {author} {\bibfnamefont {M.}~\bibnamefont {Landini}},\ }\href
  {https://arxiv.org/abs/2602.17657} {\bibinfo {title} {Realization of
  fractional fermi seas}} (\bibinfo {year} {2026}),\ \Eprint
  {https://arxiv.org/abs/2602.17657} {arXiv:2602.17657 [cond-mat.quant-gas]}
  \BibitemShut {NoStop}%
\bibitem [{\citenamefont {Yang}(1967)}]{Yang1967}%
  \BibitemOpen
  \bibfield  {author} {\bibinfo {author} {\bibfnamefont {C.~N.}\ \bibnamefont
  {Yang}},\ }\bibfield  {title} {\bibinfo {title} {Some exact results for the
  many-body problem in one dimension with repulsive delta-function
  interaction},\ }\href {https://doi.org/10.1103/PhysRevLett.19.1312}
  {\bibfield  {journal} {\bibinfo  {journal} {Phys. Rev. Lett.}\ }\textbf
  {\bibinfo {volume} {19}},\ \bibinfo {pages} {1312} (\bibinfo {year}
  {1967})}\BibitemShut {NoStop}%
\bibitem [{\citenamefont {Guan}\ \emph {et~al.}(2013)\citenamefont {Guan},
  \citenamefont {Batchelor},\ and\ \citenamefont {Lee}}]{Guan2013}%
  \BibitemOpen
  \bibfield  {author} {\bibinfo {author} {\bibfnamefont {X.-W.}\ \bibnamefont
  {Guan}}, \bibinfo {author} {\bibfnamefont {M.~T.}\ \bibnamefont
  {Batchelor}},\ and\ \bibinfo {author} {\bibfnamefont {C.}~\bibnamefont
  {Lee}},\ }\bibfield  {title} {\bibinfo {title} {Fermi gases in one dimension:
  From bethe ansatz to experiments},\ }\href
  {https://doi.org/10.1103/RevModPhys.85.1633} {\bibfield  {journal} {\bibinfo
  {journal} {Rev. Mod. Phys.}\ }\textbf {\bibinfo {volume} {85}},\ \bibinfo
  {pages} {1633} (\bibinfo {year} {2013})}\BibitemShut {NoStop}%
\bibitem [{\citenamefont {Mesty\'an}\ \emph {et~al.}(2019)\citenamefont
  {Mesty\'an}, \citenamefont {Bertini}, \citenamefont {Piroli},\ and\
  \citenamefont {Calabrese}}]{Mestyan2019}%
  \BibitemOpen
  \bibfield  {author} {\bibinfo {author} {\bibfnamefont {M.}~\bibnamefont
  {Mesty\'an}}, \bibinfo {author} {\bibfnamefont {B.}~\bibnamefont {Bertini}},
  \bibinfo {author} {\bibfnamefont {L.}~\bibnamefont {Piroli}},\ and\ \bibinfo
  {author} {\bibfnamefont {P.}~\bibnamefont {Calabrese}},\ }\bibfield  {title}
  {\bibinfo {title} {Spin-charge separation effects in the low-temperature
  transport of one-dimensional fermi gases},\ }\href
  {https://doi.org/10.1103/PhysRevB.99.014305} {\bibfield  {journal} {\bibinfo
  {journal} {Phys. Rev. B}\ }\textbf {\bibinfo {volume} {99}},\ \bibinfo
  {pages} {014305} (\bibinfo {year} {2019})}\BibitemShut {NoStop}%
\bibitem [{\citenamefont {Scopa}\ \emph {et~al.}(2021)\citenamefont {Scopa},
  \citenamefont {Calabrese},\ and\ \citenamefont {Piroli}}]{Scopa2021}%
  \BibitemOpen
  \bibfield  {author} {\bibinfo {author} {\bibfnamefont {S.}~\bibnamefont
  {Scopa}}, \bibinfo {author} {\bibfnamefont {P.}~\bibnamefont {Calabrese}},\
  and\ \bibinfo {author} {\bibfnamefont {L.}~\bibnamefont {Piroli}},\
  }\bibfield  {title} {\bibinfo {title} {Real-time spin-charge separation in
  one-dimensional fermi gases from generalized hydrodynamics},\ }\href
  {https://doi.org/10.1103/PhysRevB.104.115423} {\bibfield  {journal} {\bibinfo
   {journal} {Phys. Rev. B}\ }\textbf {\bibinfo {volume} {104}},\ \bibinfo
  {pages} {115423} (\bibinfo {year} {2021})}\BibitemShut {NoStop}%
\bibitem [{\citenamefont {Scopa}\ \emph {et~al.}(2022)\citenamefont {Scopa},
  \citenamefont {Calabrese},\ and\ \citenamefont {Piroli}}]{Scopa2022}%
  \BibitemOpen
  \bibfield  {author} {\bibinfo {author} {\bibfnamefont {S.}~\bibnamefont
  {Scopa}}, \bibinfo {author} {\bibfnamefont {P.}~\bibnamefont {Calabrese}},\
  and\ \bibinfo {author} {\bibfnamefont {L.}~\bibnamefont {Piroli}},\
  }\bibfield  {title} {\bibinfo {title} {Generalized hydrodynamics of the
  repulsive spin-$\frac{1}{2}$ fermi gas},\ }\href
  {https://doi.org/10.1103/PhysRevB.106.134314} {\bibfield  {journal} {\bibinfo
   {journal} {Phys. Rev. B}\ }\textbf {\bibinfo {volume} {106}},\ \bibinfo
  {pages} {134314} (\bibinfo {year} {2022})}\BibitemShut {NoStop}%
\bibitem [{\citenamefont {Doyon}\ and\ \citenamefont
  {Yoshimura}(2017)}]{Doyon2017}%
  \BibitemOpen
  \bibfield  {author} {\bibinfo {author} {\bibfnamefont {B.}~\bibnamefont
  {Doyon}}\ and\ \bibinfo {author} {\bibfnamefont {T.}~\bibnamefont
  {Yoshimura}},\ }\bibfield  {title} {\bibinfo {title} {{A note on generalized
  hydrodynamics: inhomogeneous fields and other concepts}},\ }\href
  {https://doi.org/10.21468/SciPostPhys.2.2.014} {\bibfield  {journal}
  {\bibinfo  {journal} {SciPost Phys.}\ }\textbf {\bibinfo {volume} {2}},\
  \bibinfo {pages} {014} (\bibinfo {year} {2017})}\BibitemShut {NoStop}%
\bibitem [{\citenamefont {Bastianello}\ \emph {et~al.}(2019)\citenamefont
  {Bastianello}, \citenamefont {Alba},\ and\ \citenamefont
  {Caux}}]{Bastianello2019}%
  \BibitemOpen
  \bibfield  {author} {\bibinfo {author} {\bibfnamefont {A.}~\bibnamefont
  {Bastianello}}, \bibinfo {author} {\bibfnamefont {V.}~\bibnamefont {Alba}},\
  and\ \bibinfo {author} {\bibfnamefont {J.-S.}\ \bibnamefont {Caux}},\
  }\bibfield  {title} {\bibinfo {title} {Generalized hydrodynamics with
  space-time inhomogeneous interactions},\ }\href
  {https://doi.org/10.1103/PhysRevLett.123.130602} {\bibfield  {journal}
  {\bibinfo  {journal} {Phys. Rev. Lett.}\ }\textbf {\bibinfo {volume} {123}},\
  \bibinfo {pages} {130602} (\bibinfo {year} {2019})}\BibitemShut {NoStop}%
\bibitem [{\citenamefont {Doyon}\ \emph {et~al.}(2018)\citenamefont {Doyon},
  \citenamefont {Yoshimura},\ and\ \citenamefont {Caux}}]{Doyon2018}%
  \BibitemOpen
  \bibfield  {author} {\bibinfo {author} {\bibfnamefont {B.}~\bibnamefont
  {Doyon}}, \bibinfo {author} {\bibfnamefont {T.}~\bibnamefont {Yoshimura}},\
  and\ \bibinfo {author} {\bibfnamefont {J.-S.}\ \bibnamefont {Caux}},\
  }\bibfield  {title} {\bibinfo {title} {Soliton gases and generalized
  hydrodynamics},\ }\href {https://doi.org/10.1103/PhysRevLett.120.045301}
  {\bibfield  {journal} {\bibinfo  {journal} {Phys. Rev. Lett.}\ }\textbf
  {\bibinfo {volume} {120}},\ \bibinfo {pages} {045301} (\bibinfo {year}
  {2018})}\BibitemShut {NoStop}%
\bibitem [{\citenamefont {Doyon}\ \emph {et~al.}(2024)\citenamefont {Doyon},
  \citenamefont {H\"ubner},\ and\ \citenamefont {Yoshimura}}]{Doyon2024TT}%
  \BibitemOpen
  \bibfield  {author} {\bibinfo {author} {\bibfnamefont {B.}~\bibnamefont
  {Doyon}}, \bibinfo {author} {\bibfnamefont {F.}~\bibnamefont {H\"ubner}},\
  and\ \bibinfo {author} {\bibfnamefont {T.}~\bibnamefont {Yoshimura}},\
  }\bibfield  {title} {\bibinfo {title} {New classical integrable systems from
  generalized $t\overline{T}$-deformations},\ }\href
  {https://doi.org/10.1103/PhysRevLett.132.251602} {\bibfield  {journal}
  {\bibinfo  {journal} {Phys. Rev. Lett.}\ }\textbf {\bibinfo {volume} {132}},\
  \bibinfo {pages} {251602} (\bibinfo {year} {2024})}\BibitemShut {NoStop}%
\bibitem [{\citenamefont {Urilyon}\ \emph {et~al.}(2026)\citenamefont
  {Urilyon}, \citenamefont {Biagetti}, \citenamefont {Kethepalli},\ and\
  \citenamefont {De~Nardis}}]{Urilyon2026}%
  \BibitemOpen
  \bibfield  {author} {\bibinfo {author} {\bibfnamefont {A.}~\bibnamefont
  {Urilyon}}, \bibinfo {author} {\bibfnamefont {L.}~\bibnamefont {Biagetti}},
  \bibinfo {author} {\bibfnamefont {J.}~\bibnamefont {Kethepalli}},\ and\
  \bibinfo {author} {\bibfnamefont {J.}~\bibnamefont {De~Nardis}},\ }\bibfield
  {title} {\bibinfo {title} {Simulating generalized fluids via interacting wave
  packet evolution},\ }\href {https://doi.org/10.1103/b587-8yyt} {\bibfield
  {journal} {\bibinfo  {journal} {Phys. Rev. B}\ }\textbf {\bibinfo {volume}
  {113}},\ \bibinfo {pages} {014314} (\bibinfo {year} {2026})}\BibitemShut
  {NoStop}%
\bibitem [{\citenamefont {Bastianello}\ and\ \citenamefont
  {De~Luca}(2019)}]{BastianelloFlux2019}%
  \BibitemOpen
  \bibfield  {author} {\bibinfo {author} {\bibfnamefont {A.}~\bibnamefont
  {Bastianello}}\ and\ \bibinfo {author} {\bibfnamefont {A.}~\bibnamefont
  {De~Luca}},\ }\bibfield  {title} {\bibinfo {title} {Integrability-protected
  adiabatic reversibility in quantum spin chains},\ }\href
  {https://doi.org/10.1103/PhysRevLett.122.240606} {\bibfield  {journal}
  {\bibinfo  {journal} {Phys. Rev. Lett.}\ }\textbf {\bibinfo {volume} {122}},\
  \bibinfo {pages} {240606} (\bibinfo {year} {2019})}\BibitemShut {NoStop}%
\bibitem [{\citenamefont {Fürst}\ \emph {et~al.}(2013)\citenamefont {Fürst},
  \citenamefont {Lukkarinen}, \citenamefont {Mei},\ and\ \citenamefont
  {Spohn}}]{Furst2013}%
  \BibitemOpen
  \bibfield  {author} {\bibinfo {author} {\bibfnamefont {M.~L.~R.}\
  \bibnamefont {Fürst}}, \bibinfo {author} {\bibfnamefont {J.}~\bibnamefont
  {Lukkarinen}}, \bibinfo {author} {\bibfnamefont {P.}~\bibnamefont {Mei}},\
  and\ \bibinfo {author} {\bibfnamefont {H.}~\bibnamefont {Spohn}},\ }\bibfield
   {title} {\bibinfo {title} {Derivation of a matrix-valued boltzmann equation
  for the hubbard model},\ }\href
  {https://doi.org/10.1088/1751-8113/46/48/485002} {\bibfield  {journal}
  {\bibinfo  {journal} {Journal of Physics A: Mathematical and Theoretical}\
  }\textbf {\bibinfo {volume} {46}},\ \bibinfo {pages} {485002} (\bibinfo
  {year} {2013})}\BibitemShut {NoStop}%
\bibitem [{\citenamefont {Bertini}\ \emph {et~al.}(2015)\citenamefont
  {Bertini}, \citenamefont {Essler}, \citenamefont {Groha},\ and\ \citenamefont
  {Robinson}}]{Bertini2015}%
  \BibitemOpen
  \bibfield  {author} {\bibinfo {author} {\bibfnamefont {B.}~\bibnamefont
  {Bertini}}, \bibinfo {author} {\bibfnamefont {F.~H.~L.}\ \bibnamefont
  {Essler}}, \bibinfo {author} {\bibfnamefont {S.}~\bibnamefont {Groha}},\ and\
  \bibinfo {author} {\bibfnamefont {N.~J.}\ \bibnamefont {Robinson}},\
  }\bibfield  {title} {\bibinfo {title} {Prethermalization and thermalization
  in models with weak integrability breaking},\ }\href
  {https://doi.org/10.1103/PhysRevLett.115.180601} {\bibfield  {journal}
  {\bibinfo  {journal} {Phys. Rev. Lett.}\ }\textbf {\bibinfo {volume} {115}},\
  \bibinfo {pages} {180601} (\bibinfo {year} {2015})}\BibitemShut {NoStop}%
\bibitem [{\citenamefont {Surace}\ and\ \citenamefont
  {Motrunich}(2023)}]{Surace2023}%
  \BibitemOpen
  \bibfield  {author} {\bibinfo {author} {\bibfnamefont {F.~M.}\ \bibnamefont
  {Surace}}\ and\ \bibinfo {author} {\bibfnamefont {O.}~\bibnamefont
  {Motrunich}},\ }\bibfield  {title} {\bibinfo {title} {Weak integrability
  breaking perturbations of integrable models},\ }\href
  {https://doi.org/10.1103/PhysRevResearch.5.043019} {\bibfield  {journal}
  {\bibinfo  {journal} {Phys. Rev. Res.}\ }\textbf {\bibinfo {volume} {5}},\
  \bibinfo {pages} {043019} (\bibinfo {year} {2023})}\BibitemShut {NoStop}%
\bibitem [{\citenamefont {M\o{}ller}\ and\ \citenamefont
  {Schmiedmayer}(2020)}]{Moller2020}%
  \BibitemOpen
  \bibfield  {author} {\bibinfo {author} {\bibfnamefont {F.~S.}\ \bibnamefont
  {M\o{}ller}}\ and\ \bibinfo {author} {\bibfnamefont {J.}~\bibnamefont
  {Schmiedmayer}},\ }\bibfield  {title} {\bibinfo {title} {{Introducing iFluid:
  a numerical framework for solving hydrodynamical equations in integrable
  models}},\ }\href {https://doi.org/10.21468/SciPostPhys.8.3.041} {\bibfield
  {journal} {\bibinfo  {journal} {SciPost Phys.}\ }\textbf {\bibinfo {volume}
  {8}},\ \bibinfo {pages} {041} (\bibinfo {year} {2020})}\BibitemShut {NoStop}%
\bibitem [{sup()}]{suppmat}%
  \BibitemOpen
  \href@noop {} {\bibinfo  {journal} {Supplementary Material for details on the
  numerical implementation of GHD equations.}\ }\BibitemShut {NoStop}%
\bibitem [{\citenamefont {Senaratne}\ \emph {et~al.}(2022)\citenamefont
  {Senaratne}, \citenamefont {Cavazos-Cavazos}, \citenamefont {Wang},
  \citenamefont {He}, \citenamefont {Chang}, \citenamefont {Kafle},
  \citenamefont {Pu}, \citenamefont {Guan},\ and\ \citenamefont
  {Hulet}}]{Senaratne2022}%
  \BibitemOpen
\bibfield  {journal} {  }\bibfield  {author} {\bibinfo {author} {\bibfnamefont
  {R.}~\bibnamefont {Senaratne}}, \bibinfo {author} {\bibfnamefont
  {D.}~\bibnamefont {Cavazos-Cavazos}}, \bibinfo {author} {\bibfnamefont
  {S.}~\bibnamefont {Wang}}, \bibinfo {author} {\bibfnamefont {F.}~\bibnamefont
  {He}}, \bibinfo {author} {\bibfnamefont {Y.-T.}\ \bibnamefont {Chang}},
  \bibinfo {author} {\bibfnamefont {A.}~\bibnamefont {Kafle}}, \bibinfo
  {author} {\bibfnamefont {H.}~\bibnamefont {Pu}}, \bibinfo {author}
  {\bibfnamefont {X.-W.}\ \bibnamefont {Guan}},\ and\ \bibinfo {author}
  {\bibfnamefont {R.~G.}\ \bibnamefont {Hulet}},\ }\bibfield  {title} {\bibinfo
  {title} {Spin-charge separation in a one-dimensional fermi gas with tunable
  interactions},\ }\href {https://doi.org/10.1126/science.abn1719} {\bibfield
  {journal} {\bibinfo  {journal} {Science}\ }\textbf {\bibinfo {volume}
  {376}},\ \bibinfo {pages} {1305} (\bibinfo {year} {2022})},\ \Eprint
  {https://arxiv.org/abs/https://www.science.org/doi/pdf/10.1126/science.abn1719}
  {https://www.science.org/doi/pdf/10.1126/science.abn1719} \BibitemShut
  {NoStop}%
\bibitem [{\citenamefont {Klauser}\ and\ \citenamefont
  {Caux}(2011)}]{Klauser2011}%
  \BibitemOpen
  \bibfield  {author} {\bibinfo {author} {\bibfnamefont {A.}~\bibnamefont
  {Klauser}}\ and\ \bibinfo {author} {\bibfnamefont {J.-S.}\ \bibnamefont
  {Caux}},\ }\bibfield  {title} {\bibinfo {title} {Equilibrium thermodynamic
  properties of interacting two-component bosons in one dimension},\ }\href
  {https://doi.org/10.1103/PhysRevA.84.033604} {\bibfield  {journal} {\bibinfo
  {journal} {Phys. Rev. A}\ }\textbf {\bibinfo {volume} {84}},\ \bibinfo
  {pages} {033604} (\bibinfo {year} {2011})}\BibitemShut {NoStop}%
\bibitem [{\citenamefont {Robinson}\ and\ \citenamefont
  {Konik}(2017)}]{Robinson2017}%
  \BibitemOpen
  \bibfield  {author} {\bibinfo {author} {\bibfnamefont {N.~J.}\ \bibnamefont
  {Robinson}}\ and\ \bibinfo {author} {\bibfnamefont {R.~M.}\ \bibnamefont
  {Konik}},\ }\bibfield  {title} {\bibinfo {title} {Excitations in the
  yang–gaudin bose gas},\ }\href {https://doi.org/10.1088/1742-5468/aa6f46}
  {\bibfield  {journal} {\bibinfo  {journal} {Journal of Statistical Mechanics:
  Theory and Experiment}\ }\textbf {\bibinfo {volume} {2017}},\ \bibinfo
  {pages} {063101} (\bibinfo {year} {2017})}\BibitemShut {NoStop}%
\bibitem [{\citenamefont {Kosevich}\ \emph {et~al.}(1998)\citenamefont
  {Kosevich}, \citenamefont {Gann}, \citenamefont {Zhukov},\ and\ \citenamefont
  {Voronov}}]{Kosevich1998}%
  \BibitemOpen
  \bibfield  {author} {\bibinfo {author} {\bibfnamefont {A.~M.}\ \bibnamefont
  {Kosevich}}, \bibinfo {author} {\bibfnamefont {V.~V.}\ \bibnamefont {Gann}},
  \bibinfo {author} {\bibfnamefont {A.~I.}\ \bibnamefont {Zhukov}},\ and\
  \bibinfo {author} {\bibfnamefont {V.~P.}\ \bibnamefont {Voronov}},\
  }\bibfield  {title} {\bibinfo {title} {Magnetic soliton motion in a
  nonuniform magnetic field},\ }\href {https://doi.org/10.1134/1.558674}
  {\bibfield  {journal} {\bibinfo  {journal} {Journal of Experimental and
  Theoretical Physics}\ }\textbf {\bibinfo {volume} {87}},\ \bibinfo {pages}
  {401} (\bibinfo {year} {1998})}\BibitemShut {NoStop}%
\bibitem [{\citenamefont {Kosevich}(2001)}]{Kosevich2001}%
  \BibitemOpen
  \bibfield  {author} {\bibinfo {author} {\bibfnamefont {A.~M.}\ \bibnamefont
  {Kosevich}},\ }\bibfield  {title} {\bibinfo {title} {Bloch oscillations of
  magnetic solitons as an example of dynamical localization of quasiparticles
  in a uniform external field (review)},\ }\href
  {https://doi.org/10.1063/1.1388415} {\bibfield  {journal} {\bibinfo
  {journal} {Low Temperature Physics}\ }\textbf {\bibinfo {volume} {27}},\
  \bibinfo {pages} {513} (\bibinfo {year} {2001})}\BibitemShut {NoStop}%
\bibitem [{\citenamefont {Congy}\ \emph {et~al.}(2016)\citenamefont {Congy},
  \citenamefont {Kamchatnov},\ and\ \citenamefont {Pavloff}}]{Congy2016}%
  \BibitemOpen
  \bibfield  {author} {\bibinfo {author} {\bibfnamefont {T.}~\bibnamefont
  {Congy}}, \bibinfo {author} {\bibfnamefont {A.~M.}\ \bibnamefont
  {Kamchatnov}},\ and\ \bibinfo {author} {\bibfnamefont {N.}~\bibnamefont
  {Pavloff}},\ }\bibfield  {title} {\bibinfo {title} {{Dispersive hydrodynamics
  of nonlinear polarization waves in two-component Bose-Einstein
  condensates}},\ }\href {https://doi.org/10.21468/SciPostPhys.1.1.006}
  {\bibfield  {journal} {\bibinfo  {journal} {SciPost Phys.}\ }\textbf
  {\bibinfo {volume} {1}},\ \bibinfo {pages} {006} (\bibinfo {year}
  {2016})}\BibitemShut {NoStop}%
\bibitem [{\citenamefont {Bresolin}\ \emph {et~al.}(2023)\citenamefont
  {Bresolin}, \citenamefont {Roy}, \citenamefont {Ferrari}, \citenamefont
  {Recati},\ and\ \citenamefont {Pavloff}}]{Bresolin2023}%
  \BibitemOpen
  \bibfield  {author} {\bibinfo {author} {\bibfnamefont {S.}~\bibnamefont
  {Bresolin}}, \bibinfo {author} {\bibfnamefont {A.}~\bibnamefont {Roy}},
  \bibinfo {author} {\bibfnamefont {G.}~\bibnamefont {Ferrari}}, \bibinfo
  {author} {\bibfnamefont {A.}~\bibnamefont {Recati}},\ and\ \bibinfo {author}
  {\bibfnamefont {N.}~\bibnamefont {Pavloff}},\ }\bibfield  {title} {\bibinfo
  {title} {Oscillating solitons and ac josephson effect in ferromagnetic
  bose-bose mixtures},\ }\href {https://doi.org/10.1103/PhysRevLett.130.220403}
  {\bibfield  {journal} {\bibinfo  {journal} {Phys. Rev. Lett.}\ }\textbf
  {\bibinfo {volume} {130}},\ \bibinfo {pages} {220403} (\bibinfo {year}
  {2023})}\BibitemShut {NoStop}%
\bibitem [{\citenamefont {Manakov}(1974)}]{manakov1974}%
  \BibitemOpen
  \bibfield  {author} {\bibinfo {author} {\bibfnamefont {S.~V.}\ \bibnamefont
  {Manakov}},\ }\bibfield  {title} {\bibinfo {title} {On the theory of
  two-dimensional stationary self-focusing of electromagnetic waves},\
  }\href@noop {} {\bibfield  {journal} {\bibinfo  {journal} {Soviet
  Physics-JETP}\ }\textbf {\bibinfo {volume} {38}},\ \bibinfo {pages} {248}
  (\bibinfo {year} {1974})}\BibitemShut {NoStop}%
\bibitem [{\citenamefont {Koch}\ \emph {et~al.}(2022)\citenamefont {Koch},
  \citenamefont {Caux},\ and\ \citenamefont {Bastianello}}]{Koch2022}%
  \BibitemOpen
  \bibfield  {author} {\bibinfo {author} {\bibfnamefont {R.}~\bibnamefont
  {Koch}}, \bibinfo {author} {\bibfnamefont {J.-S.}\ \bibnamefont {Caux}},\
  and\ \bibinfo {author} {\bibfnamefont {A.}~\bibnamefont {Bastianello}},\
  }\bibfield  {title} {\bibinfo {title} {Generalized hydrodynamics of the
  attractive non-linear schr\"odinger equation},\ }\href
  {https://doi.org/10.1088/1751-8121/ac53c3} {\bibfield  {journal} {\bibinfo
  {journal} {Journal of Physics A: Mathematical and Theoretical}\ }\textbf
  {\bibinfo {volume} {55}},\ \bibinfo {pages} {134001} (\bibinfo {year}
  {2022})}\BibitemShut {NoStop}%
\bibitem [{\citenamefont {Koch}\ and\ \citenamefont
  {Bastianello}(2023)}]{Koch2023}%
  \BibitemOpen
  \bibfield  {author} {\bibinfo {author} {\bibfnamefont {R.}~\bibnamefont
  {Koch}}\ and\ \bibinfo {author} {\bibfnamefont {A.}~\bibnamefont
  {Bastianello}},\ }\bibfield  {title} {\bibinfo {title} {{Exact thermodynamics
  and transport in the classical sine-Gordon model}},\ }\href
  {https://doi.org/10.21468/SciPostPhys.15.4.140} {\bibfield  {journal}
  {\bibinfo  {journal} {SciPost Phys.}\ }\textbf {\bibinfo {volume} {15}},\
  \bibinfo {pages} {140} (\bibinfo {year} {2023})}\BibitemShut {NoStop}%
\bibitem [{\citenamefont {Bastianello}\ \emph {et~al.}(2024)\citenamefont
  {Bastianello}, \citenamefont {Krajnik},\ and\ \citenamefont
  {Ilievski}}]{Bastianello2024}%
  \BibitemOpen
  \bibfield  {author} {\bibinfo {author} {\bibfnamefont {A.}~\bibnamefont
  {Bastianello}}, \bibinfo {author} {\bibfnamefont {i.~c.~v.}\ \bibnamefont
  {Krajnik}},\ and\ \bibinfo {author} {\bibfnamefont {E.}~\bibnamefont
  {Ilievski}},\ }\bibfield  {title} {\bibinfo {title} {Landau-lifschitz
  magnets: Exact thermodynamics and transport},\ }\href
  {https://doi.org/10.1103/PhysRevLett.133.107102} {\bibfield  {journal}
  {\bibinfo  {journal} {Phys. Rev. Lett.}\ }\textbf {\bibinfo {volume} {133}},\
  \bibinfo {pages} {107102} (\bibinfo {year} {2024})}\BibitemShut {NoStop}%
\bibitem [{\citenamefont {Lannig}\ \emph {et~al.}(2020)\citenamefont {Lannig},
  \citenamefont {Schmied}, \citenamefont {Pr\"ufer}, \citenamefont {Kunkel},
  \citenamefont {Strohmaier}, \citenamefont {Strobel}, \citenamefont
  {Gasenzer}, \citenamefont {Kevrekidis},\ and\ \citenamefont
  {Oberthaler}}]{Lannig2020}%
  \BibitemOpen
  \bibfield  {author} {\bibinfo {author} {\bibfnamefont {S.}~\bibnamefont
  {Lannig}}, \bibinfo {author} {\bibfnamefont {C.-M.}\ \bibnamefont {Schmied}},
  \bibinfo {author} {\bibfnamefont {M.}~\bibnamefont {Pr\"ufer}}, \bibinfo
  {author} {\bibfnamefont {P.}~\bibnamefont {Kunkel}}, \bibinfo {author}
  {\bibfnamefont {R.}~\bibnamefont {Strohmaier}}, \bibinfo {author}
  {\bibfnamefont {H.}~\bibnamefont {Strobel}}, \bibinfo {author} {\bibfnamefont
  {T.}~\bibnamefont {Gasenzer}}, \bibinfo {author} {\bibfnamefont {P.~G.}\
  \bibnamefont {Kevrekidis}},\ and\ \bibinfo {author} {\bibfnamefont {M.~K.}\
  \bibnamefont {Oberthaler}},\ }\bibfield  {title} {\bibinfo {title}
  {Collisions of three-component vector solitons in bose-einstein
  condensates},\ }\href {https://doi.org/10.1103/PhysRevLett.125.170401}
  {\bibfield  {journal} {\bibinfo  {journal} {Phys. Rev. Lett.}\ }\textbf
  {\bibinfo {volume} {125}},\ \bibinfo {pages} {170401} (\bibinfo {year}
  {2020})}\BibitemShut {NoStop}%
\bibitem [{\citenamefont {Cominotti}\ \emph {et~al.}(2023)\citenamefont
  {Cominotti}, \citenamefont {Berti}, \citenamefont {Dulin}, \citenamefont
  {Rogora}, \citenamefont {Lamporesi}, \citenamefont {Carusotto}, \citenamefont
  {Recati}, \citenamefont {Zenesini},\ and\ \citenamefont
  {Ferrari}}]{Cominotti2023}%
  \BibitemOpen
  \bibfield  {author} {\bibinfo {author} {\bibfnamefont {R.}~\bibnamefont
  {Cominotti}}, \bibinfo {author} {\bibfnamefont {A.}~\bibnamefont {Berti}},
  \bibinfo {author} {\bibfnamefont {C.}~\bibnamefont {Dulin}}, \bibinfo
  {author} {\bibfnamefont {C.}~\bibnamefont {Rogora}}, \bibinfo {author}
  {\bibfnamefont {G.}~\bibnamefont {Lamporesi}}, \bibinfo {author}
  {\bibfnamefont {I.}~\bibnamefont {Carusotto}}, \bibinfo {author}
  {\bibfnamefont {A.}~\bibnamefont {Recati}}, \bibinfo {author} {\bibfnamefont
  {A.}~\bibnamefont {Zenesini}},\ and\ \bibinfo {author} {\bibfnamefont
  {G.}~\bibnamefont {Ferrari}},\ }\bibfield  {title} {\bibinfo {title}
  {Ferromagnetism in an extended coherently coupled atomic superfluid},\ }\href
  {https://doi.org/10.1103/PhysRevX.13.021037} {\bibfield  {journal} {\bibinfo
  {journal} {Phys. Rev. X}\ }\textbf {\bibinfo {volume} {13}},\ \bibinfo
  {pages} {021037} (\bibinfo {year} {2023})}\BibitemShut {NoStop}%
\bibitem [{\citenamefont {Menyuk}(1989)}]{Menyuk1989}%
  \BibitemOpen
  \bibfield  {author} {\bibinfo {author} {\bibfnamefont {C.}~\bibnamefont
  {Menyuk}},\ }\bibfield  {title} {\bibinfo {title} {Pulse propagation in an
  elliptically birefringent kerr medium},\ }\href
  {https://doi.org/10.1109/3.40656} {\bibfield  {journal} {\bibinfo  {journal}
  {IEEE Journal of Quantum Electronics}\ }\textbf {\bibinfo {volume} {25}},\
  \bibinfo {pages} {2674} (\bibinfo {year} {1989})}\BibitemShut {NoStop}%
\bibitem [{\citenamefont {Kivshar}\ and\ \citenamefont
  {Agrawal}(2003)}]{kivshar2003}%
  \BibitemOpen
  \bibfield  {author} {\bibinfo {author} {\bibfnamefont {Y.~S.}\ \bibnamefont
  {Kivshar}}\ and\ \bibinfo {author} {\bibfnamefont {G.~P.}\ \bibnamefont
  {Agrawal}},\ }\href@noop {} {\emph {\bibinfo {title} {Optical solitons: from
  fibers to photonic crystals}}}\ (\bibinfo  {publisher} {Academic press},\
  \bibinfo {year} {2003})\BibitemShut {NoStop}%
\bibitem [{\citenamefont {Scopa}\ \emph {et~al.}(2026)\citenamefont {Scopa},
  \citenamefont {Zechmann}, \citenamefont {Michael}, \citenamefont
  {De~Nardis},\ and\ \citenamefont {Bastianello}}]{Zenodo}%
  \BibitemOpen
  \bibfield  {author} {\bibinfo {author} {\bibfnamefont {S.}~\bibnamefont
  {Scopa}}, \bibinfo {author} {\bibfnamefont {P.}~\bibnamefont {Zechmann}},
  \bibinfo {author} {\bibfnamefont {K.}~\bibnamefont {Michael}}, \bibinfo
  {author} {\bibfnamefont {J.}~\bibnamefont {De~Nardis}},\ and\ \bibinfo
  {author} {\bibfnamefont {A.}~\bibnamefont {Bastianello}},\ }\bibfield
  {title} {\bibinfo {title} {Generalized hydrodynamics of bloch oscillations in
  the absence of a lattice},\ }\href {https://doi.org/10.5281/zenodo.20272044}
  {10.5281/zenodo.20272044} (\bibinfo {year} {2026})\BibitemShut {NoStop}%
\end{thebibliography}%

\clearpage
\onecolumngrid
\newpage

\setcounter{equation}{0}  
\setcounter{figure}{0}
\setcounter{page}{1}
\setcounter{section}{0}    
\renewcommand\thesection{\arabic{section}}    
\renewcommand\thesubsection{\arabic{subsection}}    
\renewcommand{\thetable}{S\arabic{table}}
\renewcommand{\theequation}{S\arabic{equation}}
\renewcommand{\thefigure}{S\arabic{figure}}
\setcounter{secnumdepth}{2}  

\begin{center}
{\Large \textbf{Supplementary Material}}\\ \ \\
{\large \textbf{\titleinfo}}
\ \\ \ \\
\end{center}
\bigskip
\bigskip

This Supplementary Material covers the technical aspects about the numerical solution of the GHD equation.
For the sake of convenience, we work with dimensionless units setting $\hbar\to 1$, $m\to 1/2$ and $F=1$. A C++ implementation of the algorithm here described is available on Zenodo~\cite{Zenodo}.

\section{Discretizing the integral equations}
\label{sec_dis}

The integral equations governing the thermodynamics and hydrodynamics of the YG model are discretized as it follows. We introduce a large cutoff $\Lambda$ for the rapidity space $\lambda\in(-\infty,\infty)\to \lambda\in(-\Lambda,\Lambda)$, and a discretization $\{\lambda_i\}_{i=0}^N$ with the convention $\lambda_0=-\Lambda$ and $\lambda_N=\Lambda$. The discretization divides the rapidity space into intervals $[\lambda_i,\lambda_{i+1})$ and the occupancy and all the other relevant functions are discretized picking the middle point
\be
\vartheta_j(\lambda)\to \vartheta_{(i,j)}\equiv \vartheta_j\big(\tfrac{\lambda_{i+1}+\lambda_i}{2}\big)\, .
\ee

The $\lambda-$discretization does not need to be equally spaced, and it is actually not convenient since the occupancy and other relevant functions are smooth at large rapidities. We tried different discretization and chose the optimal one, as we later discuss. The string index is truncated to $j\in\{0,1,...,j_\text{max}\}$: the composite index $(i,j)$ is vectorized in a space of dimension $N\times (j_\text{max}+1)$.
Integral equations governing the dressing operation and thermodynamics are seemingly transformed in matrix-valued equations through proper discretizations of the associated integrals.
First, we properly regularize the scattering kernel $\phi_j(\lambda)$ to accommodate the periodic boundary conditions in the rapidity space, which would be spoiled by the sharp cutoff $\Lambda$. Hence, we define $\phi_{j;\Lambda}(\lambda)=\frac{\dd }{\dd\lambda} 2\arctan\left(\tan\left(\frac{\pi \lambda}{2\Lambda}\right)\frac{4\Lambda}{c j\pi}\right)$: the deformation $\phi_{j}(\lambda)\to \phi_{j;\Lambda}(\lambda)$ enforces the periodicity of $\lambda$ with period $2\Lambda$, and it converges to the correct value for large cutoffs $\lim_{\Lambda\to \infty}\phi_{j;\Lambda}(\lambda)=\phi_{j}(\lambda)$. Furthermore, the two kernels have the same norm $\int_{-\infty}^{\infty}\dd\lambda\, \phi_{j}(\lambda)=\int_{-\Lambda}^{\Lambda}\dd\lambda\, \phi_{j;\Lambda}(\lambda)=2\pi$.

From the deformation $\phi_{j;\Lambda}(\lambda)$, we build the deformation of the full string-string scattering kernel $\phi_{j,j'}(\lambda)\to \phi_{j,j';\Lambda}(\lambda)$ with the same definitions of the original one.
We then define the discrete kernels $\Phi_{(i,j),(i',j')}$ as
\be
\Phi_{(i,j),(i',j')}=\int_{\lambda_{i'}}^{\lambda_{i'+1}} \dd\lambda' \, \phi_{j,j';\Lambda}\left(\frac{\lambda_{i}+\lambda_{i+1}}{2}-\lambda'\right)\, .
\ee
Notice that the integral over $\lambda'$ is easily performed analytically.
The dressing equations for a test function $[\tau_{j}(\lambda)]^\dr$ are then discretized as
\be
[\tau_{j}(\lambda)]^\dr=\tau_j(\lambda)-\frac{1}{2\pi}\sum_{j'}\int \phi_{j,j'}(\lambda-\lambda')\vartheta_{j'}(\lambda')[\tau_{j'}(\lambda')]^\dr\hspace{1pc}\to\hspace{1pc} [\tau_{(i,j)}]^\dr=\tau_{(i,j)}-\sum_{(i',j')}\frac{1}{2\pi}\Phi_{(i,j),(i',j')}\vartheta_{(i',j')}[\tau_{(i',j')}]^\dr\, .
\ee
Once discretized, the dressing equations are efficiently solved with linear algebra libraries. The non-linear integral equations to determine thermal states are efficiently solved with the Newton's method.

\subsubsection{On the choice of the discretization}
The discretization is tailored to get the best possible convergence with the least number of points in the discretization, but we did not attempt a systematic optimization. The value of $j_\text{max}$ is tuned in such a way the population of higher strings is negligible: for the numerical results we showed $j_\text{max}=2$ sufficies. The value of the cutoff $\Lambda$ is chosen in such a way $\int_0^\lambda \dd\lambda'\phi_j(\lambda')=2\arctan\left(\frac{2\lambda}{c j}\right)$ is well converged to $\int_0^\lambda\dd\lambda'\,\phi_{j;\Lambda}(\lambda')=2\arctan\left(\tan\left(\frac{\pi \lambda}{2\Lambda}\right)\frac{4\Lambda}{c j\pi}\right)$ in the whole interval $\lambda\in (-\Lambda,\Lambda)$ for all the relevant values of the string index. Since the convergence worsens for larger strings and large rapidities, we use $\left|2\arctan\left(\tan\left(\frac{\pi \lambda}{2\Lambda}\right)\frac{4\Lambda}{c j\pi}\right)-2\arctan\left(\frac{2\lambda}{c j}\right)\right|_{j=2j_\text{max},\lambda=\Lambda}=\pi-2\arctan\left(\frac{\Lambda}{cj_\text{max}}\right)$ as an estimator. In our simulations we chose $\Lambda=50$.
The non-linear discretization is chosen as $\lambda_i=\Lambda f\left(\frac{i}{N}-1/2\right)/f(1)$ with $i\in \{0,1,...,N\}$ and $f(x)$ a suitable odd function. We choose $f(x)=0.5x+x^7$, as it compromised a denser discretization for smaller rapidities (where the occupancies and other functions have a non-trivial profile) and a more rarefied discretization approaching the cutoff. We attained a good convergence for the thermodynamics and forthcoming hydrodynamics with $N=100$.

\section{The Solution of the GHD equations through the method of characteristics}

The GHD equations are solved with the method of characteristics~\cite{Bastianello2019,Moller2020} in the space of the filling functions. Before discretizing, the GHD equation $\partial_t\vartheta_j(\lambda)+a^\text{eff}_j(\lambda)\partial_\lambda \vartheta_j(\lambda)=0$ can be implicitly solved as (for the sake of clarity, below we restore the explicit time dependence)
\be
\vartheta_{j; t+\dd t}(\lambda)=\vartheta_{j; t}\left(\lambda(t+\dd t)\right)\, ,
\ee
where $\lambda(t')=\lambda-\int_{t}^{t'} \dd \tau\, a^\eff_{j; \tau}(\lambda(\tau))$. This is an implicit solution, which can be used for an $O(\dd t^2)$ algorithm by posing
\be
\vartheta_{j; t+\dd t}(\lambda)\simeq\vartheta_{j; t}\left(\lambda-\dd t a^\eff_{j; t+\dd t/2}\left(\lambda-\tfrac{\dd t}{2} a^\eff_{j; t+\dd t/2}(\lambda)\right)\right)\, .\label{eq_ghd_2ord}
\ee
The filling $\vartheta_{j;t}$ is updated at time $t+\dd t$ evaluating the effective acceleration at the intermediate time, hence the algorithm needs also the filling function $\vartheta_{j;t+\dd t/2}$. The filling $\vartheta_{j;t=0}$ is initialized from the solution of thermodynamics, while $\vartheta_{j;t=\dd t/2}$ is obtained solving the GHD equation with a first-order algorithm in time with a very small time step. Then, the so-obtained two-time filling functions are alternatively evolved using Eq.~\eqref{eq_ghd_2ord}.
When passing from the continuous rapidity space in Eq.~\eqref{eq_ghd_2ord} to a properly discretized rapidity set, it is crucial employing a good discretization and interpolation. 
In practice, we experienced that a stable interpolation requires a much finer discretization that what is needed for the integral equations described in Section~\ref{sec_dis}, but further increasing the number of points in the discretization would have severely slowed down the numerical solution of the dressing equations.
Therefore, we introduce a finer rapidity discretization $\tilde{\lambda}_{\tilde{i}}$ such that the previously-introduced discretization is a subset of it $\lambda_i=\tilde{\lambda}_{i q}$ with $q$ an integer. The filling fraction $\vartheta_{j}(\lambda)$ is discretized on the finer grid giving $\tilde{\vartheta}_{(\tilde{i},j)}$ and Eq.~\eqref{eq_ghd_2ord} implemented through a second-order interpolation in the fine rapidity grid. We obtain the discretized filling to be employed in Eq.~\eqref{eq_ghd_2ord} by local averaging of the finer discretization $\vartheta_{(i,j)}=\frac{1}{\lambda_{i+1}-\lambda_i}\sum_{s=0}^{q-1}(\tilde{\lambda}_{iq+s+1}-\tilde{\lambda}_{iq+s})\tilde{\vartheta}_{(iq+s,j)}$. The effective acceleration is computed on the grid $\{\lambda_i\}_{i=0}^N$ and interpolated beyond that.
We run simulations using a factor $q=10$ between the two discretizations, which we experienced to be stable for several Bloch oscillations.
In the simulations, $\dd t$ has to be sufficiently small to ensure stability: for these discretization parameters, $\dd t=2\times 10^{-5}$ provided stable results.
The convergence of the code has been consistently checked on the conserved laws (number of particles) and by ensuring changes in the discretization were not visibly affecting data.

\section{Additional data: artificial suppression of two-magnon states}
\begin{figure}[h!]
\centering
	\includegraphics[width=0.43\columnwidth]{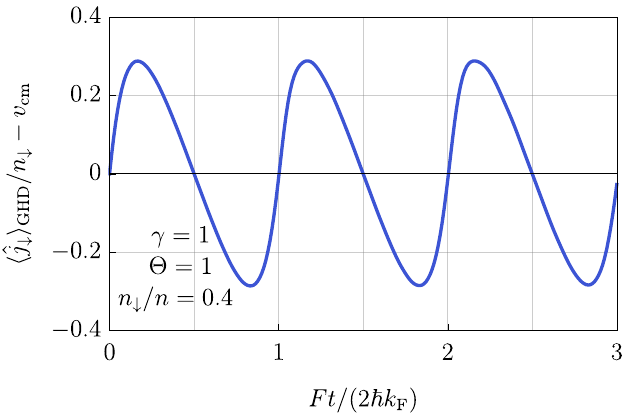}
\caption{\textbf{Perfect Bloch oscillations in the absence of two magnon excitations.---} The normalized impurity current $\langle \hat{j}_\dw\rangle_\text{GHD}/n_\dw$ is shown for an evolved thermal state with $\gamma=1$ and adimensional temperature $\Theta=1$ and large magnetization, where we artificially suppress excitations of strings $j>1$. Bloch oscillations are perfect with period $Ft/(2\hbar k_\text{F})$.}
	\label{fig_SM}
\end{figure}

In the main text, we discuss that as far as two (or higher) magnon excitations are not excited in the states, Bloch oscillations are prefect even at a finite magnetization.
To further support this claim, we consider a thermal ensemble with the same parameters of Fig. \ref{fig_4} (a), where we artificially limit the strings to $j=0$ and $j=1$, and tune the relative chemical potential to reach a large impurity density. We stress that the so-obtained state is not a Gibbs Ensemble any longer, but an exotic GGE.
Even when the impurity density is comparable with that of data shown in Fig. \ref{fig_4} (a) with an appreciable drift form perfect Bloch oscillations, the latter are re-established suppressing magnonic bound states.

\end{document}